\documentclass[aps,preprint,showpacs]{revtex4}
\usepackage{amssymb}
\usepackage{graphicx}
\usepackage{bm}
\usepackage{amsmath}

\input{tcilatex}

\begin{document}

\title{A Fast Impurity Solver Based on Gutzwiller variational approach}
\author{Jia-Ning Zhuang, Lei Wang, Zhong Fang, Xi Dai}
\date{\today}

\begin{abstract}
A fast impurity solver for the dynamical mean field theory(DMFT)
named Two Mode Approximation (TMA)   is proposed based on the
Gutzwiller variational approach, which captures the main features of
both the coherent and incoherent motion of the electrons. The new
solver works with real frequency at zero temperature and it provides
directly the spectral function of the electrons. It can be easily
generalized to multi-orbital impurity problems with general on-site
interactions, which makes it very useful in LDA+DMFT. Benchmarks on
one and two band Hubbard models are presented, and the results agree
well with those of Exact Diagonalization (ED).
\end{abstract}

\maketitle

\setcounter{MaxMatrixCols}{10} 

\setcounter{MaxMatrixCols}{10}



\bigskip

\section{Introduction}

The accurate calculation of the electronic structure of materials starting
from first principles is a challenging problem in condensed matter science.
The local density approximation (LDA) based on density functional theory
(DFT) is a widely used \textit{ab initio} method \cite{L(S)DA}, which has
been successfully applied to study the properties of simple metals and
semiconductors as well as the band insulators. However, it can not be
applied to those materials containing partially filled narrow bands from
\textit{d} or \textit{f} shells, because of the so called strong correlation
effect.

In LDA the wave like nature rather than the atomic feature of the
electronic state is emphasized, so it is more suitable to describe
those wide energy bands contributed by the electrons from outer
shells. While for the electrons from those unclosed inner shells
like 3\textit{d} or 5\textit{f} shells, some atomic features such as
the multiplet structure remain, which are poorly described by LDA.
Therefore for those strongly correlated materials, we have to
implement LDA with some many-body techniques which can deal with the
strong correlation effect and capture most of the atomic features.

One notable example of the first-principle schemes is the LDA+U method \cite%
{LDA+U}, which can successfully describe many interesting effects such as
spin, orbital and charge ordering in transition metal compounds \cite{LDA+U
application}. Although LDA+U can capture the static orbital and spin
dependent physics quite well, it still can not consider the dynamical
correlation effect, which causes lots of interesting phenomena like Mott
transition \cite{Mott1} \cite{Mott2} \cite{Mott3}.

Another attempt is to use Gutzwiller variational approach \cite%
{Gutzwiller_old} \cite{He3} to take into account the correlation effect
(LDA+G), which is superior to LDA+U and has been successfully applied to
many systems\cite{Gutzwiller} \cite{xydeng} \cite{xydeng_long}. LDA+G
treatment has its advantage in describing ground state and low energy
excited states, but it can not properly describe the finite temperature and
dynamical properties due to the lack of high energy excited states. In order
to capture the overall features of a correlated materials, more
sophisticated approaches are needed.

During the past twenty years, the dynamical mean field theory (DMFT) \cite%
{DMFT} has been quickly developed to be a powerful method to solve
the strongly correlated models on the lattice. DMFT maps the lattice
models to the corresponding quantum impurity models subject to
self-consistency conditions. Unlike the normal static mean field
approaches, DMFT keeps the full local dynamics induced by the local
interaction. DMFT has been successfully applied to various of
correlation problems, such as the Mott transition in Hubbard model
\cite{mit1} \cite{mit2}, the pseudo gap behavior in high $T_c$
cuperates \cite{pseudo_gap} and the heavy fermion system
\cite{qimiao} \cite{qimiao2}. Since DMFT can capture quite
accurately the correlation feature induced by the on-site Coulomb
interaction and LDA can take care of the periodic potential as well
as the long range part of the Coulomb interaction, the combination
of the two methods should be a very useful scheme for the first
priciple calculation of correlation materials. In the past twenty
years, LDA+DMFT has been developed very quickly and successfully
applied to many systems\cite{LDA+DMFT}, see \cite{apply1}
\cite{apply2} \cite{Kotliar LDA+DMFT} and \cite{held_review} for
reviews of the recent developments and applications.

In LDA+DMFT, one encounters the problem of how to efficiently solve
quantum impurity problems with self- consistently determined bath
degrees of freedom. A fast impurity solver can be regarded as the
\textit{engine} of DMFT, which determines the efficiency and
accuracy of DMFT. Many impurity solvers have been developed in the
past twenty years, which can be divided into analytical methods and
numerical methods. The analytical mthods include
equation of motion (EOM) method \cite{eom}, Hubbard-I approximation \cite%
{Savrasov-Hub1} \cite{Savrasov-Hub1_2}, iterative perturbation theory (IPT)
\cite{IPT1} \cite{IPT2}, the Non-crossing approximation(NCA) \cite{NCA} and
the fluctuation exchange approximation(FLEX) \cite{FLEX}. And the numerical
methods include exact diagonalization (ED) \cite{ED} , Hirsch-Fye Quantum
Monte Carlo methods \cite{QMC} \cite{Hirsh-Fye} and the numerical
renormalization group (NRG) \cite{NRG}. Most recently a powerful
continuous-time quantum Monte Carlo (CTQMC) solver \cite{CTQMC} \cite%
{weak_ctqmc} has also been developed and applied to several realistic
matterials\cite{Pu_Gabi} \cite{LaOFeAs_Gabi}.

All these impurity solvers have their own advantages and the limitations as
well. Since most of the novel quantum phenomena in condensed matter physics
happen in very low temperature, it is always very important for us to study
the low temperature properties of the correlated materials using LDA+DMFT.
Up to now, the impurity solvers which can work at extremely low temperature
are ED, IPT and NRG. Among them, IPT can only apply to the single band
system, ED and NRG are numerically quite heavy for a general multi-band
system. Therefore it is very useful to develop an impurity solver working at
zero temperature, which satisfies the following criteria. i) It can capture
both the low energy quasi-particle physics and the high energy Hubbard
bands. ii) It works with real frequency and gives the real time dynamical
properties directly. iii) It is easy to be generalized to realistic
multi-band systems.

Here we propose a fast impurity solver based on Gutzwiller
variational approach\cite{Gutzwiller} which has the above three
advantages. Gutzwiller variational wave function associated with
Gutzwiller approximation was first proposed to deal with lattice
problems such as the
Hubbard model and the periodical Anderson model\cite{anderson_lattice1} \cite%
{anderson_lattice2}. In the present paper, we apply a generalized
Gutzwiller method called Two Mode Approximation (TMA) to calculate
the Green's function for a quantum impurity model generated by DMFT.
TMA is first proposed in reference\cite{qianghua} to calculate the
spectral function for the lattice mode. Here we generalize it to the
quantum impurity problem and make it a useful impurity solver for
DMFT.

In TMA three different types of variational wave functions are
constructed for the ground states, low energy quasi-particle states
and high energy excited state respectively. All the variational
parameters appearing in different wave functions are determined by
minimizing the ground state energy, based on which we can obtain the
electronic spectral functions over the full frequency range. The
computational time is mainly determined by the minimization of the
ground state energy and is similar with the previous study on
lattice problem\cite{osmt_sbmf}, which can be easily done even on a
single PC. This makes the present approach a fast general solver for
LDA+DMFT studies.

The paper is organized as follows. In section II we give the
derivation of the method and prove that the sum rule for the
electronic spectral function is satisfied. In Section III we
benchmark our new impurity solver on the two-band Hubbard model with
DMFT+ED. Finally a summary and the conclusions are made in section
IV.


\bigskip

\section{Derivation of the method}

\subsection{Gutzwiller ground state}

Let us first consider the following multi-orbital impurity Hamiltonian

\begin{eqnarray*}
\hat{H}_{imp} &=&\hat{H}_{band}+\hat{H}_{local}+\hat{H}_{V} \\
\hat{H}_{band} &=&\sum_{k\sigma }\epsilon _{k\sigma }\hat{c}_{k\sigma }^{+}%
\hat{c}_{k\sigma } \\
\hat{H}_{local} &=&\sum_{\sigma ,\sigma ^{\prime }}U_{\sigma \sigma ^{\prime
}}\hat{n}_{f\sigma }\hat{n}_{f\sigma ^{\prime }}+\sum_{\sigma }\varepsilon
_{\sigma }\hat{n}_{f\sigma } \\
\hat{H}_{V} &=&\sum_{k\sigma }V_{k\sigma }(\hat{c}_{k\sigma }^{+}\hat{f}%
_{\sigma }+h.c.)\label{Himp}
\end{eqnarray*}%
where $k$ denotes the energy levels in the bath and $\sigma$ is the joint index for
orbital and spin. In Gutzwiller variational approach, the ground state of
the above Hamiltonian can be written as

\begin{equation}
|\Psi \rangle =\hat{P}|0\rangle  \label{GW_ground}
\end{equation}%
Where $\hat{P}$ is the Gutzwiller projector and $|0\rangle $
is a single Slatter Determinant like wave function. Both of  $\hat{P}$ and $|0\rangle $ will be
determined by minimizing the ground state energy. Following reference \cite%
{Gutzwiller}, the Gutzwiller projector can be written in terms of
the projection operators of the atomic eigen states as
\begin{equation}
\hat{P}=\sum_{\Gamma }\frac{\sqrt{m_{\Gamma }}}{\sqrt{m_{\Gamma }^{0}}}\hat{m%
}_{\Gamma }  \label{GW_projector}
\end{equation}

In equation (\ref{GW_projector}), the operator $\hat{m}_{\Gamma }\equiv
|\Gamma \rangle \langle \Gamma |$ is the projector to the eigen
states $|\Gamma \rangle $ of the  atomic Hamiltonian $\hat{H}_{local}$, and $m_{\Gamma }$ are the variational parameters
introduced in the Gutzwiller theory. Note that if $\hat{H}_{local}$ only
contains density-density interactions, the atomic eigen states are known as
the Fock states as the following\cite{Gutzwiller},

\begin{eqnarray}
\Gamma &\in &\{\varnothing ;(1),...,(2N);(1,2),(2,3),...(2N-1,2N)  \notag \\
&&;...(1,..,2N)\}  \label{local basis}
\end{eqnarray},where $N$ is the number of orbitals.
$m_{\Gamma }^{0}$ is defined as

\begin{eqnarray}
m_{\Gamma }^{0} &\equiv &\langle 0|\hat{m}_{\Gamma }|0\rangle  \label{mgamma}
\end{eqnarray}%
Using the operator equalities
\begin{eqnarray}
\hat{m}_{\Gamma } &=&\prod\limits_{\sigma \in \Gamma }\hat{n}_{f\sigma
}\prod\limits_{\sigma \bar{\in}\Gamma }(1-\hat{n}_{f\sigma }) \label{GW_decompose}\\
\hat{n}_{f\sigma } &=&\sum_{\Gamma \ni \sigma }\hat{m}_{\Gamma }
\label{GW_decompose2}
\end{eqnarray}%
with the definition $n_{f\sigma }^{0}\equiv \langle 0|\hat{n}_{f\sigma
}|0\rangle $ and $n_{f\sigma }\equiv \langle \Psi |\hat{n}_{f\sigma }|\Psi
\rangle $, one can prove that $m_{\Gamma }^{0}=\prod\limits_{\sigma \in
\Gamma }n_{f\sigma }^{0}\prod\limits_{\sigma \bar{\in}\Gamma }(1-n_{f\sigma
}^{0})$, $n_{f\sigma }^{0}=\sum_{\Gamma \ni \sigma }m_{\Gamma }^{0}$ and $%
n_{f\sigma }=\sum_{\Gamma \ni \sigma }m_{\Gamma }$. We would emphasize that $%
n_{f\sigma }^{0}=n_{f\sigma }$ for Gutzwiller type wave functions with pure
density-density interaction, which greatly simplify the computation\cite%
{Gutzwiller,xydeng}.Therefore the Gutzwiller ground state energy of this impurity model reads

\begin{equation}
E_{g}=\frac{\langle 0|\hat{P}\hat{H}_{imp}\hat{P}|0\rangle }{\langle 0|\hat{P%
}^{2}|0\rangle }  \label{Gutzwiller ground state}
\end{equation}%
the denominator can be expressed as
\begin{equation*}
\langle 0|\hat{P}^{2}|0\rangle =\sum_{\Gamma }m_{\Gamma }=1
\end{equation*}%
while the numerator can be calculated by decomposing the projectors
as in equation (\ref{GW_decompose}) and applying the Wick's
theorem\cite{wick}. Finally we obtain the ground state energy as

\begin{eqnarray*}
E_{g} &=&\sum_{k\sigma }\epsilon _{k\sigma }\langle 0|\hat{c}_{k\sigma }^{+}%
\hat{c}_{k\sigma }|0\rangle +\sum_{\Gamma }E_{\Gamma }m_{\Gamma } \\
&+&\sum_{k\sigma }z_{\sigma }V_{k\sigma }\langle 0|\hat{c}_{k\sigma }^{+}%
\hat{f}_{\sigma }+h.c.|0\rangle
\end{eqnarray*}%
with

\begin{equation*}
z_{\sigma }=\sum_{\Gamma \ni \sigma ,\Gamma ^{\prime }=\Gamma \setminus
\sigma }\frac{\sqrt{m_{\Gamma }m_{\Gamma ^{\prime }}}}{\sqrt{n_{f\sigma
}^{0}(1-n_{f\sigma }^{0})}}
\end{equation*}%
The ground state wave function $|\Psi \rangle $ can be obtained by
minimizing the above energy functional respect to the $m_{\Gamma }$
and non-interacting wave function $|0\rangle
$\cite{Gutzwiller,xydeng} along with the following constraints.
\begin{eqnarray}
\sum_{\Gamma }m_{\Gamma } &=&1 \\
n_{f\sigma } &=&\sum_{\Gamma \ni \sigma }m_{\Gamma }
\end{eqnarray}

\bigskip

\subsection{\textbf{zero-temperature Green's function}}

For the impurity Hamiltonian Eq.(\ref{Himp}), the retarded Green's function
for the electrons on the impurity site reads
\begin{eqnarray}
G_{\sigma }^{imp}(\omega +i\eta ) &=&\sum_{n}\frac{\langle \Psi |\hat{f}%
_{\sigma }|n\rangle \langle n|\hat{f}_{\sigma }^{\dagger }|\Psi \rangle }{%
\omega +i\eta -E_{n}+E_{g}}  \notag \\
&&+\sum_{m}\frac{\langle \Psi |\hat{f}_{\sigma }^{\dagger }|m\rangle \langle
m|\hat{f}_{\sigma }|\Psi \rangle }{\omega +i\eta +E_{m}-E_{g}}  \label{Gimp}
\end{eqnarray}%

where $|\Psi \rangle $ is the ground state of $\hat{H}_{imp\text{ }}$with
the eigen energy $E_{g}$, $|n\rangle $ ($|m\rangle $)are the eigenstates of $%
\hat{H}_{imp\text{ }}$ with one more (less) electron than the ground state. $%
E_{n}$ and $E_{m}$ are the corresponding eigenvalues.  The above expression
is exact if the summation of $n$ and $m$ includes all the eigenstates. In
the present paper, we apply the two mode approximation (TMA) to solve the quantum
impurity problem, in which we  limit the above
summation in a truncated Hilbert space formed by finite number of
excited states over the Gutzwiller variational ground state \cite%
{Gunnarsson,qianghua}. In order to capture the basic feature of the
electronic spectral function efficiently, we have to include two
types of excited states in TMA, namely the quasi-particle
excitations which give the right Fermi liquid behavior in low
energy, and the high energy excited states which are responsible for
the Hubbard bands or the atomic multiplet features. The former are
called quasi-particle states and the latter are called bare-particle
states in the present paper\cite{qianghua}. The ansatz for the
excited states are the following,
\begin{eqnarray*}
|+k\sigma \rangle &=&\hat{c}_{k\sigma }^{\dagger }\hat{P}|0\rangle \\
|UHB\rangle &=&\hat{f}_{\sigma }^{\dagger }\hat{P}|0\rangle \\
|QE\rangle &=&\hat{P}\hat{f}_{\sigma }^{\dagger }|0\rangle \\
|-k\sigma \rangle &=&\hat{c}_{k\sigma }\hat{P}|0\rangle \\
|LHB\rangle &=&\hat{f}_{\sigma }\hat{P}|0\rangle \\
|QH\rangle \ &=&\hat{P}\hat{f}_{\sigma }|0\rangle
\end{eqnarray*}%
where $|QE\rangle$ ($|QH\rangle $) are the quasi-particle (quasi-hole)
states, $|UHB\rangle $ ($|LHB\rangle$ ) are the bare-particle (bare-hole)
states, and $|+- k\sigma \rangle$ represent the excitations in the bath.

The excited states listed above are neither orthogonal nor normalized, thus
we have to calculate the overlaps $\mathcal{O}_{\alpha \beta }\equiv \langle
\alpha |\beta \rangle \mathcal{\ }$and the matrix elements of the
Hamiltonian $\mathcal{H}_{\alpha \beta }\equiv \langle \alpha |\hat{H}|\beta
\rangle $ in this truncated Hilbert space. This procedure could be easily
done by applying Wick's theorem. We list all the necessary matrix elements
and overlaps in the Appendix.

In order to evaluate the Green's function using expression
(\ref{Gimp}), we have to first obtain the eigen states and eigen
values by solving the following generalized eigen equation in the
truncated Hilbert space.
\begin{equation*}
\mathcal{H}|l\rangle =E_{l}\mathcal{O}|l\rangle
\end{equation*}

Therefore $|l\rangle $ form a complete basis for the truncated Hilbert space
and the completeness condition $\sum_{l}|l\rangle \langle l|=1$ is satisfied
within the truncated Hilbert space. Since both the states $\hat{f}_{\sigma
}^{\dagger }\hat{P}|0\rangle$ and $\hat{f}_{\sigma }\hat{P}|0\rangle$ are
fully included in the contained Hilbert space, it is easy to prove that

\begin{eqnarray*}
(-\frac{1}{\pi })\mathrm{Im}[G_{\sigma }^{imp}(\omega +i\eta )] &=&\langle
\Psi |\hat{f}_{\sigma }\hat{f}_{\sigma }^{\dagger }+\hat{f}_{\sigma
}^{\dagger }\hat{f}_{\sigma }|\Psi \rangle \\
&=&1
\end{eqnarray*}%
, which is the sum rule of the impurity Green's function.

\section{Benchmark}

\subsection{Impurity Spectral function}

\begin{figure}[!ht]
\includegraphics[height=6cm,width=7cm,bb=15 11 459 374]{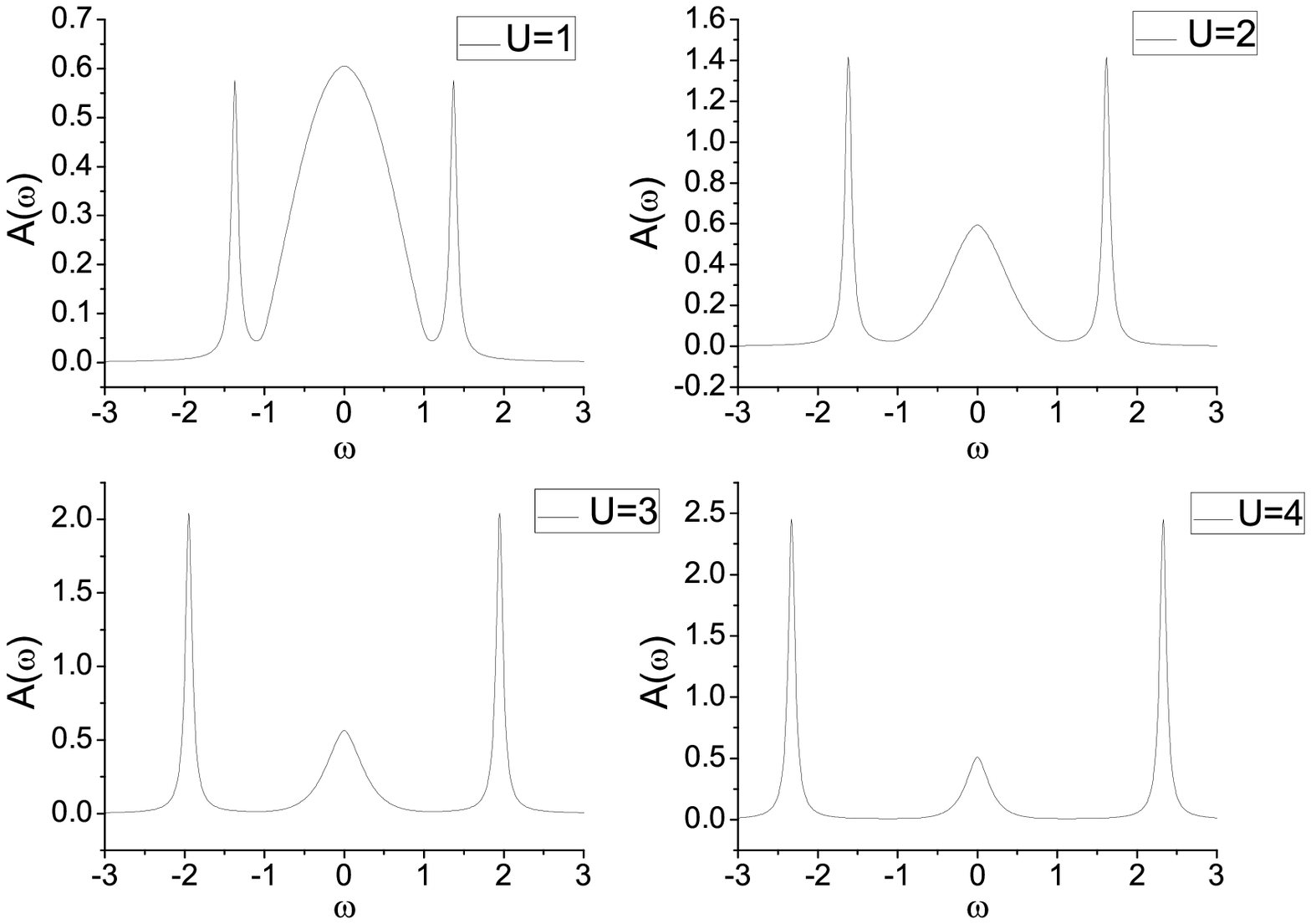}
\caption{The spectral function of electrons on the impurity site for
an single orbital impurity model with different $U$ and
semi-circular density of states in the bath.} \label{GFimp}
\end{figure}

First of all we check the spectral function obtained by TMA for a
single orbital impurity model with particle-hole symmetry. The
density of states for the heat bath is chosen to be the semicircle
with the half-width $D=1$. The spectral functions for the electron
on the impurity site with different Hubbard interaction $U$ are
shown in Fig.(\ref{GFimp}).

From Fig.(\ref{GFimp}) we find that the spectral function contains
three parts, the quasi-particle peak and two Hubbard bands. With the
increment of $U$, the spectral weight transfers from the low energy
quasi-particle part to the Hubbard bands. And in large $U$ limit,
the distance between two Hubbard bands approaches $U$. All these
features are consistent with the previous
studies on the symmetric Anderson model \cite{Hewson_book}. In Fig.(\ref%
{GFimp,psVSmf}), we compare one spectral function for an Anderson
impurity
model obtained by TMA with that by the normal Gutzwiller Approximation (GA)%
\cite{Gutzwiller_old,Gutzwiller} for the lattice model, which only contains
the quasi-particle part as

\begin{equation}
G_{imp}^{GWMF}(\omega +i\eta )=\frac{z^{2}}{\omega +i\eta +\tilde\mu
-z^{2}\Delta (\omega +i\eta )}
\end{equation}
.

Compared with normal Gutzwiller approximation (GA lattice), it is
very clear that TMA can reproduce very nicely the low energy
quasi-particle part with slightly smaller spectral weight. Therefore
the current solver can be viewed as the normal Gutzwiller
approximation implemented with the Hubbard bands in the high energy
part of the electronic spectral functions describing the atomic
features.

\subsection{Used as the impurity solver in DMFT}

The present impurity solver can be used in the dynamical mean field
theory to study the lattice models. In this paper we have studied
both the single-band and two-band Hubbard model at paramagnetic
phase with arbitrary fillings.

\subsubsection{Single-band Hubbard model}

We start with the single band Hubbard model on the Bethe lattice
with half band width $D=1$. First we check the half filling case. We
show the spectral function  with the increment of $U$ in
Fig.(\ref{spect,sb,ntot=1}), from which we see that the height of
quasi-particle peak changes little before Mott transition, but the
integral of the quasi-particle spectrum reduces as $U$ increases.
This feature is consistent with the previous results obtained by
DMFT+IPT\cite{DMFT}.

We show the results for the systems away from half filling in Fig.(\ref%
{U=2,filling varies}).

With the increment of filling factor from $N_{tot}=0.2$ to half filling $%
N_{tot}=1.0$, the spectral weight continuously transfers from the low energy
quasi-particle part to the high energy Hubbard bands, which is consistent
with the common understanding that the strong correlation effect is less
pronounced when the system is doped away from half filling.

\begin{figure}[th]
\includegraphics[height=6cm,width=9cm,bb=9 21 298
218]{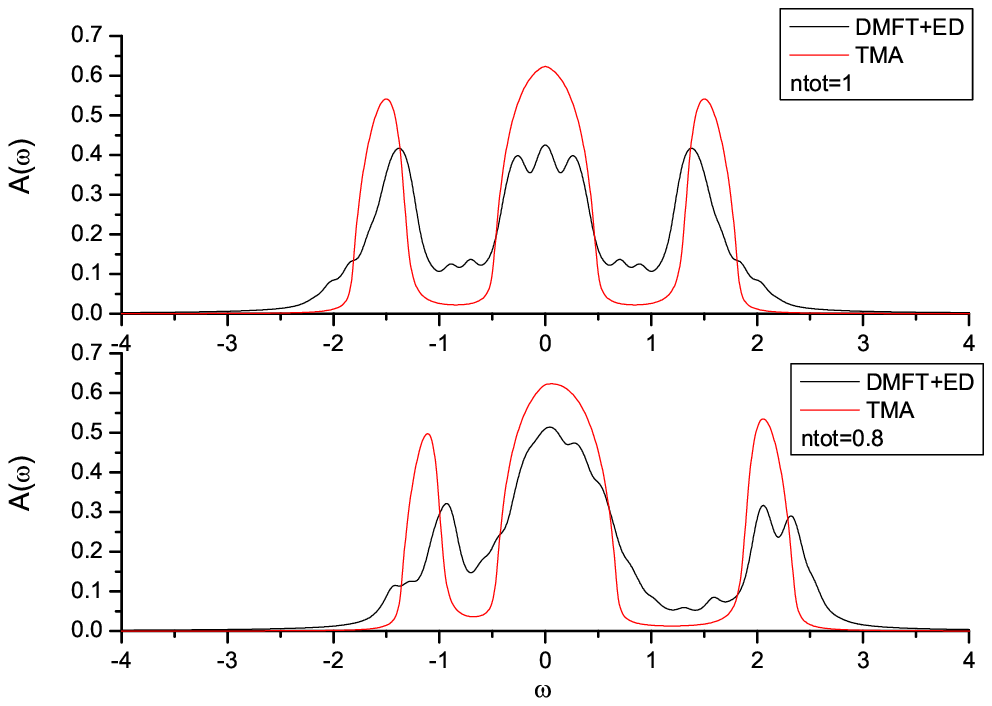} \caption{Comparison of the
DOS obtained by DMFT+TMA and DMFT+ED for single band Hubbard model
with $U=2$.} \label{onebandDOScomparewithDMFTED}
\end{figure}

In Fig.(\ref{onebandDOScomparewithDMFTED}), we quantitatively
compare the density of states (DOS) obtained by DMFT+TMA with that
by DMFT+ED. We find quite good agreement between them for both the
half filling and non-half filling cases. While we also find two
disagreements. Compared with the DMFT+ED results, the total spectral
weight of the quasi-particle part is over-estimated while the width
of the Hubbard bands is under-estimated by DMFT+TMA.

We have also calculated the quasi-particle weight $z$, which is a
characteristic quantity describing the strength of the correlation effect
and is defined as:
\begin{equation}
z_{\sigma }=(1-\frac{\partial\mathrm{Re}[\Sigma _{\sigma }(\omega
+i\eta )]}{\partial \omega })^{-1}|_{\omega =0}
\end{equation}
In Fig.(\ref{oneband,zU}) we show quasi-particle weight obtained by DMFT+TMA
as the function of $U$ for different filling factors. In the
half filling case, the value of $z$ decreases as the increment of $U$ until
the critical $U_c$ for the Mott transition. As shown in Fig.(\ref{oneband,zU}%
), $U_c$ obtained by DMFT+TMA is around $3.6$,which is bigger
than $U_{c2}=2.9$ obtained by DMFT+ED.

In Fig.(\ref{1band,z_compare}), we compare the $z$-factors obtained
by DMFT+TMA, Gutzwiller approximation on the lattice model (lattice
GA) and DMFT+ED. As discussed in reference\cite{xydeng} and
\cite{qianghua}, we can only obtain the ground state energy quite
accurately by lattice GA, but not for the $z$-factor. The reason is
quite obvious that in the lattice GA only the low energy
quasi-particle states in equation(\ref{Gimp}) can be considered,
which limits the accuracy of $z$-factor. While in TMA, we first
apply the DMFT scheme to treat the inter-site correlation on a mean
field level, which is in principle similar with GA. Then in solving
the effective impurity model, we enlarge the variational space by
including more excited states, which gives us more accurate
description of the low energy excited states and
reduces the disagreement in $z$-factor with DMFT+ED results as shown in Fig.%
(\ref{1band,z_compare}).

\subsubsection{Two-band Hubbard model on the Bethe lattice}

The situation becomes more complicated when we consider two-band models. We
start with the simplest case that the two bands are degenerate with half
bandwidth $D_1=D_2=1$ and the local part of the Hamiltonian has SU(4)
symmetry, which can be written as
\begin{eqnarray}
\hat{H}_{at} &=&U\sum_{b}\hat{n}_{b,\uparrow }\hat{n}_{b,\downarrow
}+U\sum_{\sigma ,\sigma ^{\prime }}\hat{n}_{1,\sigma }\hat{n}_{2,\sigma
^{\prime }}
\end{eqnarray}

We first show the quasi-particle weight obtained by DMFT+TMA versus $U$ at different filling
factors in Fig(\ref{twoband,zU}) and the comparison with DMFT+ED and lattice GA in Fig(\ref%
{twoband,zcompare}).

The Mott transition at integer fillings can be observed with $U_c$
slightly larger than the DMFT+ED results. As shown in
Fig(\ref{twoband,zcompare}), the improvement of the quasi-particle
weight against the lattice GA is quite dramatic, which indicates that even
for the low energy quasi-particle part the DMFT+TMA is better than
applying the GA directly to the lattice model.

The behavior of $z$ as the function of the filling factor for fixed $%
U=5.0$ is shown in Fig(\ref{twoband,U=5,zN}), from which we can find that
compared with lattice GA the results obtained by DMFT+TMA is much closer to DMFT+ED.


Next we take the Hund's coupling constant $J$ into account. Then the atomic Hamiltonian becomes%
\begin{eqnarray}
\hat{H}_{at} &=&U\sum_{b}\hat{n}_{b,\uparrow }\hat{n}_{b,\downarrow
}+U^{\prime }\sum_{\sigma ,\sigma ^{\prime }}\hat{n}_{1,\sigma }\hat{n}%
_{2,\sigma ^{\prime }}-J\sum_{\sigma }\hat{n}_{1,\sigma }\hat{n}_{2,\sigma }
\notag \\
&&+J\sum_{\sigma }\hat{c}_{1,\sigma }^{+}\hat{c}_{2,-\sigma }^{+}\hat{c}%
_{1.-\sigma }\hat{c}_{2,\sigma }+J(\hat{c}_{1,\uparrow }^{+}\hat{c}%
_{1,\downarrow }^{+}\hat{c}_{2.\downarrow }\hat{c}_{2,\uparrow }  \notag \\
&&+\hat{c}_{2,\uparrow }^{+}\hat{c}_{2,\downarrow }^{+}\hat{c}_{1.\downarrow
}\hat{c}_{1,\uparrow })
\end{eqnarray}

We have the relation $U-U^{\prime }=2J $ for system with cubic symmetry\cite%
{U-U'=2J}. In the current study, we only keep the longitudinal part
of the Hund's rule coupling and neglect the spin flip and pair
hopping terms which correspond to the last two terms in the above
equation. The results for the full rotational invariance interaction
will be studied in detail and published elsewhere.

\begin{figure}[th]
\includegraphics[height=6cm,width=8cm,bb=36 33 464 403]{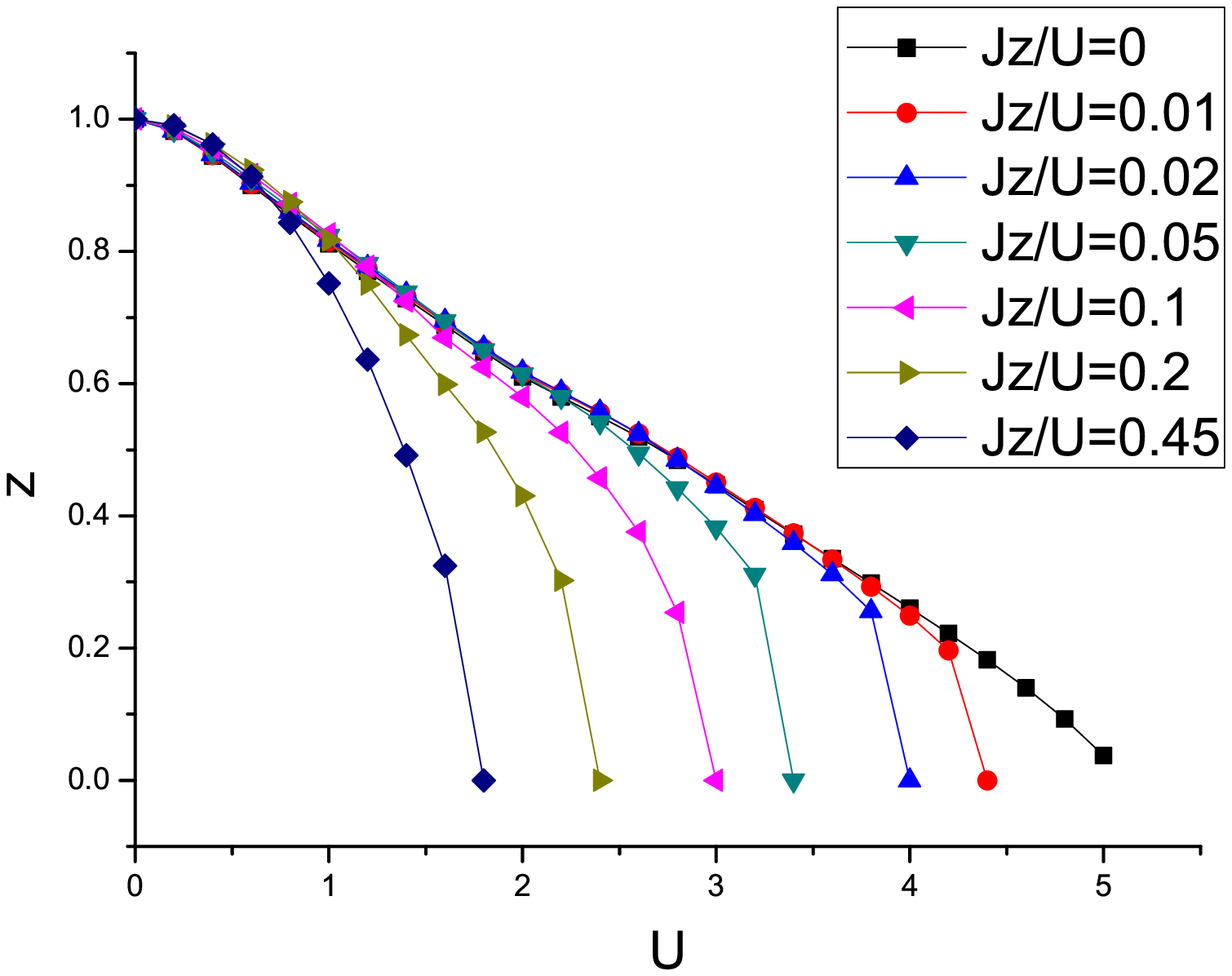}
\caption{Quasi-particle weight $z$ as the function of $U$ for the
two-degenerate-band Hubbard model with longitudinal Hund's coupling
$J_z$ obtained by DMFT+TMA at different $J_z/U$.} \label{JZoverU}
\end{figure}

The quasi-particle weight obtained by DMFT+TMA as the function of $U$ is shown in
Fig.(\ref{JZoverU}). We also compare the results with DMFT+ED in
Fig.(\ref{JZoverU,comparewithED}), from which we find that $U_c$ obtained
from TMA is larger than that of DMFT+ED as for the single band model.

In Fig.(\ref{JZoverU}), we find that the Brinkman-Rice(BR)
transition is continuous \textit{only} at the point $J_z=0$ and
first order like for all
non-zero $J_z$, which is similar with the results in reference \cite%
{Gutzwiller} obtained by rotational invariant Gutzwiller
approximation. This similarity indicates that for degenerate
multi-band Hubbard model the basic feature of the BR transition does
not strongly relies on the variational invariant treatment of the
interaction. Moreover, the similar discontinuity and the tendency
that the critical $U_c$ decreases as $J_z/U$ increases is also
obtained in \cite{Kogaarxiv}, where the self-energy functional
method is used.

However, for the non-degenerate multi-band models, i.e. the two-band
model with different band widths, the correct variational invariant
treatment is necessary to  obtain some of the qualitative features
like the orbital selective Mott transition (OSMT)\cite{Koga}.  The
detailed study for the OSMT using the variational invariant TMA
solver will be presented elsewhere. Here we only give the results
for an extreme case, where the band width
difference of the two bands  is very large. In Fig.(\ref%
{twoband,t1=0.5,t2=3,JZoverU=0.3}) and (\ref%
{twoband,t1=0.5,t2=3,JZoverU=0.3,z}), we represent the DOS as well as
 the quasi-particle weight as the functional of $U$
with fixed $J_z/U=0.3$ and half band width $D_1=1.0$, $D_2=6.0$.
Obviously in such extreme case, the system is in the orbital
selective Mott phase which is consistent with reference
\cite{OSMT_Georges}.


\section{Conclusions}

In this paper we present a new impurity solver named Two Mode
Approximation (TMA) for the multi-orbital quantum impurity model
generated by DMFT. By constructing the trial wave functions based on
the Gutzwiller variational theory not only for the ground state but
also the low energy and high energy excited states, we can obtain
the spectral functions of the electrons on the impurity level with
the satisfactory of  the sum rule. Compared with other popular
impurity solvers, TMA works with the real frequency and can obtain
both the low energy quasi-particle and high energy Hubbard band
behavior. Moreover TMA can be generalized to treat the problem with
quite general on-site interaction, which make it a good solver to be
used in LDA+DMFT.


ACKNOWLEDGEMENT: The authors would thank Q. M. Liu, X. Y. Deng, Y. Wan
and N.H. Tong for their helpful discussions. We acknowledge the  supports
from NSF of China , and that from the
973 program of China (No.2007CB925000).

\section{Appendix: Overlaps and Hamiltonian Elements}

\subsection{Overlaps}

Define

\begin{equation*}
z_{\sigma }=\sum_{\Gamma \ni \sigma ,\Gamma ^{\prime }=\Gamma \setminus
\sigma }\frac{\sqrt{m_{\Gamma }m_{\Gamma ^{\prime }}}}{\sqrt{n_{f\sigma
}^{0}(1-n_{f\sigma }^{0})}}
\end{equation*}%
the non-vanishing overlaps are

\begin{equation*}
\langle +k_{1}\sigma \left\vert +k_{2}\sigma \right\rangle =\langle 0|\hat{c}%
_{k_{1}\sigma }\hat{c}_{k_{2}\sigma }^{+}\left\vert 0\right\rangle
\end{equation*}

\begin{equation*}
\langle +k\sigma \left\vert UHB\right\rangle =z_{\sigma }\langle 0|\hat{c}%
_{k\sigma }\hat{f}_{\sigma }^{\dag }\left\vert 0\right\rangle
\end{equation*}

\begin{equation*}
\langle +k\sigma \left\vert QE\right\rangle =\langle 0|\hat{c}_{k\sigma }%
\hat{f}_{\sigma }^{\dag }\left\vert 0\right\rangle
\end{equation*}%
\qquad

\begin{equation*}
\langle UHB|UHB\rangle =(1-n_{f\sigma }^{0})
\end{equation*}

\begin{equation*}
\langle UHB|QE\rangle =z_{\sigma }(1-n_{f\sigma }^{0})
\end{equation*}

\begin{equation*}
\langle QE|QE\rangle =(1-n_{f\sigma }^{0})
\end{equation*}

\begin{equation*}
\langle -k_{1}\sigma \left\vert -k_{2}\sigma \right\rangle =\langle 0|\hat{c}%
_{k_{1}\sigma }^{+}\hat{c}_{k_{2}\sigma }\left\vert 0\right\rangle
\end{equation*}

\begin{equation*}
\langle -k\sigma \left\vert LHB\right\rangle =z_{\sigma }\langle 0|\hat{c}%
_{k\sigma }^{+}\hat{f}_{\sigma }\left\vert 0\right\rangle
\end{equation*}

\begin{equation*}
\langle -k\sigma \left\vert QH\right\rangle =\langle 0|\hat{c}_{k\sigma }^{+}%
\hat{f}_{\sigma }\left\vert 0\right\rangle
\end{equation*}

\begin{equation*}
\langle LHB|LHB\rangle =n_{f\sigma }^{0}
\end{equation*}

\begin{equation*}
\langle LHB|QH\rangle =z_{\sigma }n_{f\sigma }^{0}
\end{equation*}

\begin{equation*}
\langle LHB|LHB\rangle =n_{f\sigma }^{0}
\end{equation*}

\bigskip

\subsection{Hamiltonian Elements}

\qquad

\begin{eqnarray*}
\hat{H} &=&\hat{H}_{band}+\hat{H}_{local}+\hat{H}_{V} \\
\hat{H}_{band} &=&\sum_{k\sigma }\epsilon _{k\sigma }\hat{c}_{k\sigma }^{+}%
\hat{c}_{k\sigma } \\
\hat{H}_{local} &=&\sum_{\Gamma }E_{\Gamma }\hat{m}_{\Gamma }+\sum_{\sigma
}\varepsilon _{\sigma }\sum_{\Gamma \ni \sigma }\hat{m}_{\Gamma } \\
\hat{H}_{V} &=&\sum_{k\sigma }V_{k\sigma }(\hat{c}_{k\sigma }^{+}\hat{f}%
_{\sigma }+h.c.)
\end{eqnarray*}

\subsubsection{H\_band}

Define

\begin{eqnarray*}
x_{\sigma \sigma ^{\prime }} &=&\sum_{\substack{ \Gamma _{2}\ni \sigma
,\Gamma _{2}\ni \sigma ^{\prime }  \\ \Gamma _{1}=\Gamma _{2}\setminus
\sigma ^{\prime }}}\frac{\sqrt{m_{\Gamma _{1}}m_{\Gamma _{2}}}}{n_{f\sigma
}^{0}\sqrt{n_{f\sigma ^{\prime }}^{0}(1-n_{f\sigma ^{\prime }}^{0})}} \\
y_{\sigma \sigma ^{\prime }} &=&\sum_{\substack{ \Gamma _{1}\ni \sigma
,\Gamma _{1}\bar{\ni}\sigma ^{\prime }  \\ \Gamma _{2}=\Gamma _{1}\cup
\sigma ^{\prime }\setminus \sigma }}\frac{\sqrt{m_{\Gamma _{1}}m_{\Gamma
_{2}}}}{\sqrt{n_{f\sigma }^{0}(1-n_{f\sigma }^{0})n_{f\sigma ^{\prime
}}^{0}(1-n_{f\sigma ^{\prime }}^{0})}} \\
w_{\sigma \sigma ^{\prime }} &=&\sum_{\substack{ \Gamma _{2}\bar{\ni}\sigma
,\Gamma _{2}\bar{\ni}\sigma ^{\prime }  \\ \Gamma _{1}=\Gamma _{2}\cup
\sigma \cup \sigma ^{\prime }}}\frac{\sqrt{m_{\Gamma _{1}}m_{\Gamma _{2}}}}{%
\sqrt{n_{f\sigma }^{0}(1-n_{f\sigma }^{0})n_{f\sigma ^{\prime
}}^{0}(1-n_{f\sigma ^{\prime }}^{0})}} \\
v_{\sigma \sigma ^{\prime }} &=&\sum_{\substack{ \Gamma _{1}\bar{\ni}\sigma
,\Gamma _{1}\bar{\ni}\sigma  \\ \Gamma _{2}=\Gamma _{1}\cup \sigma ^{\prime
} }}\frac{\sqrt{m_{\Gamma _{1}}m_{\Gamma _{2}}}}{(1-n_{f\sigma }^{0})\sqrt{%
n_{f\sigma ^{\prime }}^{0}(1-n_{f\sigma ^{\prime }}^{0})}}
\end{eqnarray*}%
and

\begin{eqnarray*}
B_{\sigma \sigma ^{\prime }}^{++} &=&\sum_{\Gamma \ni \sigma ,\Gamma \ni
\sigma ^{\prime }}\frac{m_{\Gamma }}{n_{f\sigma }^{0}n_{f\sigma ^{\prime
}}^{0}} \\
B_{\sigma \sigma ^{\prime }}^{+-} &=&\sum_{\Gamma \ni \sigma ,\Gamma \bar{\ni%
}\sigma ^{\prime }}\frac{m_{\Gamma }}{n_{f\sigma }^{0}(1-n_{f\sigma ^{\prime
}}^{0})} \\
B_{\sigma \sigma ^{\prime }}^{--} &=&\sum_{\Gamma \bar{\ni}\sigma ,\Gamma
\bar{\ni}\sigma ^{\prime }}\frac{m_{\Gamma }}{(1-n_{f\sigma
}^{0})(1-n_{f\sigma ^{\prime }}^{0})}
\end{eqnarray*}
we will have

\begin{eqnarray*}
&\langle +k_{1}\sigma \left\vert \hat{H}_{band}|+k_{2}\sigma \right\rangle
=&\sum_{k^{\prime }\sigma ^{\prime }}\epsilon _{k^{\prime }\sigma ^{\prime
}}[\delta _{\sigma \sigma ^{\prime }}\langle 0|\hat{c}_{k_{1}\sigma }\hat{c}%
_{k^{\prime }\sigma }^{+}\hat{c}_{k^{\prime }\sigma }\hat{c}_{k_{2}\sigma
}^{+}\left\vert 0\right\rangle \\
&&+(1-\delta _{\sigma \sigma ^{\prime }}) \\
&&\times (B_{\sigma \sigma ^{\prime }}^{++}\langle 0|\hat{c}_{k_{1}\sigma }%
\hat{c}_{k_{2}\sigma }^{+}\hat{f}_{\sigma }^{\dag }\hat{f}_{\sigma
}\left\vert 0\right\rangle \langle 0|\hat{c}_{k^{\prime }\sigma ^{\prime
}}^{+}\hat{c}_{k^{\prime }\sigma ^{\prime }}\hat{f}_{\sigma ^{\prime
}}^{\dag }\hat{f}_{\sigma ^{\prime }}\left\vert 0\right\rangle \\
&&+B_{\sigma \sigma ^{\prime }}^{+-}\langle 0|\hat{c}_{k_{1}\sigma }\hat{c}%
_{k_{2}\sigma }^{+}\hat{f}_{\sigma }^{\dag }\hat{f}_{\sigma }\left\vert
0\right\rangle \langle 0|\hat{c}_{k^{\prime }\sigma ^{\prime }}^{+}\hat{c}%
_{k^{\prime }\sigma ^{\prime }}\hat{f}_{\sigma ^{\prime }}\hat{f}_{\sigma
^{\prime }}^{\dag }\left\vert 0\right\rangle \\
&&+B_{\sigma ^{\prime }\sigma }^{+-}\langle 0|\hat{c}_{k_{1}\sigma }\hat{c}%
_{k_{2}\sigma }^{+}\hat{f}_{\sigma }\hat{f}_{\sigma }^{\dag }\left\vert
0\right\rangle \langle 0|\hat{c}_{k^{\prime }\sigma ^{\prime }}^{+}\hat{c}%
_{k^{\prime }\sigma ^{\prime }}\hat{f}_{\sigma ^{\prime }}^{\dag }\hat{f}%
_{\sigma ^{\prime }}\left\vert 0\right\rangle \\
&&+B_{\sigma \sigma ^{\prime }}^{--}\langle 0|\hat{c}_{k_{1}\sigma }\hat{c}%
_{k_{2}\sigma }^{+}\hat{f}_{\sigma }\hat{f}_{\sigma }^{\dag }\left\vert
0\right\rangle \langle 0|\hat{c}_{k^{\prime }\sigma ^{\prime }}^{+}\hat{c}%
_{k^{\prime }\sigma ^{\prime }}\hat{f}_{\sigma ^{\prime }}\hat{f}_{\sigma
^{\prime }}^{\dag }\left\vert 0\right\rangle )]
\end{eqnarray*}

\begin{eqnarray*}
\langle +k\sigma \left\vert \hat{H}_{band}|UHB\right\rangle
&=&\sum_{k^{\prime }\sigma ^{\prime }}\epsilon _{k^{\prime }\sigma ^{\prime
}}[\delta _{\sigma \sigma ^{\prime }}z_{\sigma }\langle 0|\hat{c}_{k\sigma }%
\hat{c}_{k^{\prime }\sigma }^{+}\hat{c}_{k^{\prime }\sigma }\hat{f}_{\sigma
}^{\dag }\left\vert 0\right\rangle \\
&&+(1-\delta _{\sigma \sigma ^{\prime }})\langle 0|\hat{c}_{k\sigma }\hat{f}%
_{\sigma }^{\dag }\left\vert 0\right\rangle (x_{\sigma ^{\prime }\sigma
}\langle 0|\hat{c}_{k^{\prime }\sigma ^{\prime }}^{+}\hat{c}_{k^{\prime
}\sigma ^{\prime }}\hat{f}_{\sigma ^{\prime }}^{\dag }\hat{f}_{\sigma
^{\prime }}\left\vert 0\right\rangle +v_{\sigma ^{\prime }\sigma }\langle 0|%
\hat{c}_{k^{\prime }\sigma ^{\prime }}^{+}\hat{c}_{k^{\prime }\sigma
^{\prime }}\hat{f}_{\sigma ^{\prime }}\hat{f}_{\sigma ^{\prime }}^{\dag
}\left\vert 0\right\rangle )]
\end{eqnarray*}

\begin{eqnarray*}
\langle +k\sigma |\hat{H}_{band}|QE\rangle &=&\sum_{k^{\prime }\sigma
^{\prime }}\epsilon _{k^{\prime }\sigma ^{\prime }}[\delta _{\sigma \sigma
^{\prime }}\langle 0|\hat{c}_{k\sigma }\hat{c}_{k^{\prime }\sigma }^{+}\hat{c%
}_{k^{\prime }\sigma }\hat{f}_{\sigma }^{\dag }\left\vert 0\right\rangle \\
&&+(1-\delta _{\sigma \sigma ^{\prime }})\langle 0|\hat{c}_{k\sigma }\hat{f}%
_{\sigma }^{\dag }\left\vert 0\right\rangle (B_{\sigma \sigma ^{\prime
}}^{++}\langle 0|\hat{c}_{k^{\prime }\sigma ^{\prime }}^{+}\hat{c}%
_{k^{\prime }\sigma ^{\prime }}\hat{f}_{\sigma ^{\prime }}^{\dag }\hat{f}%
_{\sigma ^{\prime }}\left\vert 0\right\rangle +B_{\sigma \sigma ^{\prime
}}^{+-}\langle 0|\hat{c}_{k^{\prime }\sigma ^{\prime }}^{+}\hat{c}%
_{k^{\prime }\sigma ^{\prime }}\hat{f}_{\sigma ^{\prime }}\hat{f}_{\sigma
^{\prime }}^{\dag }\left\vert 0\right\rangle )]
\end{eqnarray*}

\begin{eqnarray*}
\langle UHB|\hat{H}_{band}|UHB\rangle &=&\sum_{k^{\prime }\sigma ^{\prime
}}\epsilon _{k^{\prime }\sigma ^{\prime }}[\delta _{\sigma \sigma ^{\prime
}}\langle 0|\hat{f}_{\sigma }\hat{c}_{k^{\prime }\sigma }^{+}\hat{c}%
_{k^{\prime }\sigma }\hat{f}_{\sigma }^{\dag }\left\vert 0\right\rangle \\
&&+(1-\delta _{\sigma \sigma ^{\prime }})(1-n_{f\sigma }^{0}) \\
&&\times (B_{\sigma ^{\prime }\sigma }^{+-}\langle 0|\hat{c}_{k^{\prime
}\sigma ^{\prime }}^{+}\hat{c}_{k^{\prime }\sigma ^{\prime }}\hat{f}_{\sigma
^{\prime }}^{\dag }\hat{f}_{\sigma ^{\prime }}\left\vert 0\right\rangle
+B_{\sigma \sigma ^{\prime }}^{--}\langle 0|\hat{c}_{k^{\prime }\sigma
^{\prime }}^{+}\hat{c}_{k^{\prime }\sigma ^{\prime }}\hat{f}_{\sigma
^{\prime }}\hat{f}_{\sigma ^{\prime }}^{\dag }\left\vert 0\right\rangle )
\end{eqnarray*}

\begin{eqnarray*}
\langle UHB|\hat{H}_{band}|QE\rangle &=&\sum_{k^{\prime }\sigma ^{\prime
}}\epsilon _{k^{\prime }\sigma ^{\prime }}[\delta _{\sigma \sigma ^{\prime
}}z_{\sigma }\langle 0|\hat{f}_{\sigma }\hat{c}_{k^{\prime }\sigma }^{+}\hat{%
c}_{k^{\prime }\sigma }\hat{f}_{\sigma }^{\dag }\left\vert 0\right\rangle \\
&&+(1-\delta _{\sigma \sigma ^{\prime }})(1-n_{f\sigma }^{0}) \\
&&\times (x_{\sigma ^{\prime }\sigma }\langle 0|\hat{c}_{k^{\prime }\sigma
^{\prime }}^{+}\hat{c}_{k^{\prime }\sigma ^{\prime }}\hat{f}_{\sigma
^{\prime }}^{\dag }\hat{f}_{\sigma ^{\prime }}\left\vert 0\right\rangle
+v_{\sigma ^{\prime }\sigma }\langle 0|\hat{c}_{k^{\prime }\sigma ^{\prime
}}^{+}\hat{c}_{k^{\prime }\sigma ^{\prime }}\hat{f}_{\sigma ^{\prime }}\hat{f%
}_{\sigma ^{\prime }}^{\dag }\left\vert 0\right\rangle )
\end{eqnarray*}

\begin{eqnarray*}
\langle QE|\hat{H}_{band}|QE\rangle &=&\sum_{k^{\prime }\sigma ^{\prime
}}\epsilon _{k^{\prime }\sigma ^{\prime }}[\delta _{\sigma \sigma ^{\prime
}}\langle 0|\hat{f}_{\sigma }\hat{c}_{k^{\prime }\sigma }^{+}\hat{c}%
_{k^{\prime }\sigma }\hat{f}_{\sigma }^{\dag }\left\vert 0\right\rangle \\
&&+(1-\delta _{\sigma \sigma ^{\prime }})(1-n_{f\sigma }^{0}) \\
&&\times (B_{\sigma \sigma ^{\prime }}^{++}\langle 0|\hat{c}_{k^{\prime
}\sigma ^{\prime }}^{+}\hat{c}_{k^{\prime }\sigma ^{\prime }}\hat{f}_{\sigma
^{\prime }}^{\dag }\hat{f}_{\sigma ^{\prime }}\left\vert 0\right\rangle
+B_{\sigma \sigma ^{\prime }}^{+-}\langle 0|\hat{c}_{k^{\prime }\sigma
^{\prime }}^{+}\hat{c}_{k^{\prime }\sigma ^{\prime }}\hat{f}_{\sigma
^{\prime }}\hat{f}_{\sigma ^{\prime }}^{\dag }\left\vert 0\right\rangle )
\end{eqnarray*}

\begin{eqnarray*}
\langle -k_{1}\sigma \left\vert \hat{H}_{band}|-k_{2}\sigma \right\rangle
&=&\sum_{k^{\prime }\sigma ^{\prime }}\epsilon _{k^{\prime }\sigma ^{\prime
}}[\delta _{\sigma \sigma ^{\prime }}\langle 0|\hat{c}_{k_{1}\sigma }^{+}%
\hat{c}_{k^{\prime }\sigma }^{+}\hat{c}_{k^{\prime }\sigma }\hat{c}%
_{k_{2}\sigma }\left\vert 0\right\rangle \\
&&+(1-\delta _{\sigma \sigma ^{\prime }}) \\
&&\times (B_{\sigma \sigma ^{\prime }}^{++}\langle 0|\hat{c}_{k_{1}\sigma
}^{+}\hat{c}_{k_{2}\sigma }\hat{f}_{\sigma }^{\dag }\hat{f}_{\sigma
}\left\vert 0\right\rangle \langle 0|\hat{c}_{k^{\prime }\sigma ^{\prime
}}^{+}\hat{c}_{k^{\prime }\sigma ^{\prime }}\hat{f}_{\sigma ^{\prime
}}^{\dag }\hat{f}_{\sigma ^{\prime }}\left\vert 0\right\rangle \\
&&+B_{\sigma \sigma ^{\prime }}^{+-}\langle 0|\hat{c}_{k_{1}\sigma }^{+}\hat{%
c}_{k_{2}\sigma }\hat{f}_{\sigma }^{\dag }\hat{f}_{\sigma }\left\vert
0\right\rangle \langle 0|\hat{c}_{k^{\prime }\sigma ^{\prime }}^{+}\hat{c}%
_{k^{\prime }\sigma ^{\prime }}\hat{f}_{\sigma ^{\prime }}\hat{f}_{\sigma
^{\prime }}^{\dag }\left\vert 0\right\rangle \\
&&+B_{\sigma ^{\prime }\sigma }^{+-}\langle 0|\hat{c}_{k_{1}\sigma }^{+}\hat{%
c}_{k_{2}\sigma }\hat{f}_{\sigma }\hat{f}_{\sigma }^{\dag }\left\vert
0\right\rangle \langle 0|\hat{c}_{k^{\prime }\sigma ^{\prime }}^{+}\hat{c}%
_{k^{\prime }\sigma ^{\prime }}\hat{f}_{\sigma ^{\prime }}^{\dag }\hat{f}%
_{\sigma ^{\prime }}\left\vert 0\right\rangle \\
&&+B_{\sigma \sigma ^{\prime }}^{--}\langle 0|\hat{c}_{k_{1}\sigma }^{+}\hat{%
c}_{k_{2}\sigma }\hat{f}_{\sigma }\hat{f}_{\sigma }^{\dag }\left\vert
0\right\rangle \langle 0|\hat{c}_{k^{\prime }\sigma ^{\prime }}^{+}\hat{c}%
_{k^{\prime }\sigma ^{\prime }}\hat{f}_{\sigma ^{\prime }}\hat{f}_{\sigma
^{\prime }}^{\dag }\left\vert 0\right\rangle )]
\end{eqnarray*}

\begin{eqnarray*}
\langle -k\sigma \left\vert \hat{H}_{band}|LHB\right\rangle
&=&\sum_{k^{\prime }\sigma ^{\prime }}\epsilon _{k^{\prime }\sigma ^{\prime
}}(\delta _{\sigma \sigma ^{\prime }}z_{\sigma }\langle 0|\hat{c}_{k\sigma
}^{+}\hat{c}_{k^{\prime }\sigma }^{+}\hat{c}_{k^{\prime }\sigma }\hat{f}%
_{\sigma }\left\vert 0\right\rangle \\
&&+(1-\delta _{\sigma \sigma ^{\prime }})\langle 0|\hat{c}_{k\sigma }^{+}%
\hat{f}_{\sigma }\left\vert 0\right\rangle (x_{\sigma ^{\prime }\sigma
}\langle 0|\hat{c}_{k^{\prime }\sigma ^{\prime }}^{+}\hat{c}_{k^{\prime
}\sigma ^{\prime }}\hat{f}_{\sigma ^{\prime }}^{\dag }\hat{f}_{\sigma
^{\prime }}\left\vert 0\right\rangle +v_{\sigma ^{\prime }\sigma }\langle 0|%
\hat{c}_{k^{\prime }\sigma ^{\prime }}^{+}\hat{c}_{k^{\prime }\sigma
^{\prime }}\hat{f}_{\sigma ^{\prime }}\hat{f}_{\sigma ^{\prime }}^{\dag
}\left\vert 0\right\rangle ))
\end{eqnarray*}

\begin{eqnarray*}
\langle -k\sigma \left\vert \hat{H}_{band}|QH\right\rangle
&=&\sum_{k^{\prime }\sigma ^{\prime }}\epsilon _{k^{\prime }\sigma ^{\prime
}}(\delta _{\sigma \sigma ^{\prime }}\langle 0|\hat{c}_{k\sigma }^{+}\hat{c}%
_{k^{\prime }\sigma }^{+}\hat{c}_{k^{\prime }\sigma }\hat{f}_{\sigma
}\left\vert 0\right\rangle \\
&&+(1-\delta _{\sigma \sigma ^{\prime }})\langle 0|\hat{c}_{k\sigma }^{+}%
\hat{f}_{\sigma }\left\vert 0\right\rangle (B_{\sigma \sigma ^{\prime
}}^{-+}\langle 0|\hat{c}_{k^{\prime }\sigma ^{\prime }}^{+}\hat{c}%
_{k^{\prime }\sigma ^{\prime }}\hat{f}_{\sigma ^{\prime }}^{\dag }\hat{f}%
_{\sigma ^{\prime }}\left\vert 0\right\rangle +B_{\sigma \sigma ^{\prime
}}^{--}\langle 0|\hat{c}_{k^{\prime }\sigma ^{\prime }}^{+}\hat{c}%
_{k^{\prime }\sigma ^{\prime }}\hat{f}_{\sigma ^{\prime }}\hat{f}_{\sigma
^{\prime }}^{\dag }\left\vert 0\right\rangle ))
\end{eqnarray*}

\begin{eqnarray*}
\langle LHB|\hat{H}_{band}|LHB\rangle &=&\sum_{k^{\prime }\sigma ^{\prime
}}\epsilon _{k^{\prime }\sigma ^{\prime }}[\delta _{\sigma \sigma ^{\prime
}}\langle 0|\hat{f}_{\sigma }^{\dag }\hat{c}_{k^{\prime }\sigma }^{+}\hat{c}%
_{k^{\prime }\sigma }\hat{f}_{\sigma }\left\vert 0\right\rangle \\
&&+(1-\delta _{\sigma \sigma ^{\prime }})n_{f\sigma }^{0} \\
&&\times (B_{\sigma \sigma ^{\prime }}^{++}\langle 0|\hat{c}_{k^{\prime
}\sigma ^{\prime }}^{+}\hat{c}_{k^{\prime }\sigma ^{\prime }}\hat{f}_{\sigma
^{\prime }}^{\dag }\hat{f}_{\sigma ^{\prime }}\left\vert 0\right\rangle
+B_{\sigma \sigma ^{\prime }}^{+-}\langle 0|\hat{c}_{k^{\prime }\sigma
^{\prime }}^{+}\hat{c}_{k^{\prime }\sigma ^{\prime }}\hat{f}_{\sigma
^{\prime }}\hat{f}_{\sigma ^{\prime }}^{\dag }\left\vert 0\right\rangle )
\end{eqnarray*}

\begin{eqnarray*}
\langle LHB|\hat{H}_{band}|QH\rangle &=&\sum_{k^{\prime }\sigma ^{\prime
}}\epsilon _{k^{\prime }\sigma ^{\prime }}[\delta _{\sigma \sigma ^{\prime
}}z_{\sigma }\langle 0|\hat{f}_{\sigma }^{\dag }\hat{c}_{k^{\prime }\sigma
}^{+}\hat{c}_{k^{\prime }\sigma }\hat{f}_{\sigma }\left\vert 0\right\rangle
\\
&&+(1-\delta _{\sigma \sigma ^{\prime }})n_{f\sigma }^{0} \\
&&\times (x_{\sigma ^{\prime }\sigma }\langle 0|\hat{c}_{k^{\prime }\sigma
^{\prime }}^{+}\hat{c}_{k^{\prime }\sigma ^{\prime }}\hat{f}_{\sigma
^{\prime }}^{\dag }\hat{f}_{\sigma ^{\prime }}\left\vert 0\right\rangle
+v_{\sigma ^{\prime }\sigma }\langle 0|\hat{c}_{k^{\prime }\sigma ^{\prime
}}^{+}\hat{c}_{k^{\prime }\sigma ^{\prime }}\hat{f}_{\sigma ^{\prime }}\hat{f%
}_{\sigma ^{\prime }}^{\dag }\left\vert 0\right\rangle )
\end{eqnarray*}

\begin{eqnarray*}
\langle QH|\hat{H}_{band}|QH\rangle &=&\sum_{k^{\prime }\sigma ^{\prime
}}\epsilon _{k^{\prime }\sigma ^{\prime }}[\delta _{\sigma \sigma ^{\prime
}}\langle 0|\hat{f}_{\sigma }^{\dag }\hat{c}_{k^{\prime }\sigma }^{+}\hat{c}%
_{k^{\prime }\sigma }\hat{f}_{\sigma }\left\vert 0\right\rangle \\
&&+(1-\delta _{\sigma \sigma ^{\prime }})n_{f\sigma }^{0} \\
&&\times (B_{\sigma \sigma ^{\prime }}^{-+}\langle 0|\hat{c}_{k^{\prime
}\sigma ^{\prime }}^{+}\hat{c}_{k^{\prime }\sigma ^{\prime }}\hat{f}_{\sigma
^{\prime }}^{\dag }\hat{f}_{\sigma ^{\prime }}\left\vert 0\right\rangle
+B_{\sigma \sigma ^{\prime }}^{--}\langle 0|\hat{c}_{k^{\prime }\sigma
^{\prime }}^{+}\hat{c}_{k^{\prime }\sigma ^{\prime }}\hat{f}_{\sigma
^{\prime }}\hat{f}_{\sigma ^{\prime }}^{\dag }\left\vert 0\right\rangle )
\end{eqnarray*}

\bigskip

\subsubsection{H\_local \ \ }

Here we define a function for \textit{set:}

\begin{equation*}
A_{\sigma ,\Gamma }=\{%
\begin{array}{cc}
1, & \text{\textit{if} }\sigma \in \Gamma \\
0, & \text{\textit{if} }\sigma \notin \Gamma%
\end{array}%
\end{equation*}%
then define

\begin{equation*}
S_{\Gamma }=E_{\Gamma }+\sum_{\sigma ^{\prime }}\varepsilon _{\sigma
^{\prime }}A_{\sigma ^{\prime },\Gamma }
\end{equation*}%
and

\bigskip
\begin{eqnarray*}
S_{1} &=&\sum_{\Gamma }E_{\Gamma }m_{\Gamma }+\sum_{\sigma ^{\prime
}}\varepsilon _{\sigma ^{\prime }}\sum_{\Gamma \ni \sigma ^{\prime
}}m_{\Gamma } \\
&=&\sum_{\Gamma }m_{\Gamma }S_{\Gamma }
\end{eqnarray*}%
\begin{eqnarray*}
S_{2}(\sigma ) &=&\sum_{\Gamma }E_{\Gamma }A_{\sigma ,\Gamma }\sqrt{%
m_{\Gamma }m_{\Gamma \setminus \sigma }}+\sum_{\sigma ^{\prime }}\varepsilon
_{\sigma ^{\prime }}\sum_{\Gamma \ni \sigma ^{\prime }}A_{\sigma ,\Gamma }%
\sqrt{m_{\Gamma }m_{\Gamma \setminus \sigma }} \\
&=&\sum_{\Gamma }A_{\sigma ,\Gamma }\sqrt{m_{\Gamma }m_{\Gamma \setminus
\sigma }}S_{\Gamma }
\end{eqnarray*}%
\begin{eqnarray*}
S_{3}(\sigma ) &=&\sum_{\Gamma }E_{\Gamma }A_{\sigma ,\Gamma }m_{\Gamma
}+\sum_{\sigma ^{\prime }}\varepsilon _{\sigma ^{\prime }}\sum_{\Gamma \ni
\sigma ^{\prime }}A_{\sigma ,\Gamma }m_{\Gamma } \\
&=&\sum_{\Gamma }A_{\sigma ,\Gamma }m_{\Gamma }S_{\Gamma }
\end{eqnarray*}%
\begin{eqnarray*}
S_{4}(\sigma ) &=&\sum_{\Gamma }E_{\Gamma }A_{\sigma ,\Gamma }m_{\Gamma
\setminus \sigma }+\sum_{\sigma ^{\prime }}\varepsilon _{\sigma ^{\prime
}}\sum_{\Gamma \ni \sigma ^{\prime }}A_{\sigma ,\Gamma }m_{\Gamma \setminus
\sigma } \\
&=&\sum_{\Gamma }A_{\sigma ,\Gamma }m_{\Gamma \setminus \sigma }S_{\Gamma }
\end{eqnarray*}%
\begin{eqnarray*}
S_{5}(\sigma ) &=&\sum_{\Gamma }E_{\Gamma }(1-A_{\sigma ,\Gamma })\sqrt{%
m_{\Gamma }m_{\Gamma \cup \sigma }}+\sum_{\sigma ^{\prime }}\varepsilon
_{\sigma ^{\prime }}\sum_{\Gamma \ni \sigma ^{\prime }}(1-A_{\sigma ,\Gamma
})\sqrt{m_{\Gamma }m_{\Gamma \cup \sigma }} \\
&=&\sum_{\Gamma }(1-A_{\sigma ,\Gamma })\sqrt{m_{\Gamma }m_{\Gamma \cup
\sigma }}S_{\Gamma }
\end{eqnarray*}%
\begin{eqnarray*}
S_{6}(\sigma ) &=&\sum_{\Gamma }E_{\Gamma }(1-A_{\sigma ,\Gamma })m_{\Gamma
}+\sum_{\sigma ^{\prime }}\varepsilon _{\sigma ^{\prime }}\sum_{\Gamma \ni
\sigma ^{\prime }}(1-A_{\sigma ,\Gamma })m_{\Gamma } \\
&=&\sum_{\Gamma }(1-A_{\sigma ,\Gamma })m_{\Gamma }S_{\Gamma }
\end{eqnarray*}%
\begin{eqnarray*}
S_{7}(\sigma ) &=&\sum_{\Gamma }E_{\Gamma }(1-A_{\sigma ,\Gamma })m_{\Gamma
\cup \sigma }+\sum_{\sigma ^{\prime }}\varepsilon _{\sigma ^{\prime
}}\sum_{\Gamma \ni \sigma ^{\prime }}(1-A_{\sigma ,\Gamma })m_{\Gamma \cup
\sigma } \\
&=&\sum_{\Gamma }(1-A_{\sigma ,\Gamma })m_{\Gamma \cup \sigma }S_{\Gamma }
\end{eqnarray*}%
\begin{eqnarray*}
S_{25}(\sigma ) &=&\sum_{\Gamma }E_{\Gamma }[A_{\sigma ,\Gamma }\frac{%
m_{\Gamma }}{n_{f\sigma }^{0}}-(1-A_{\sigma ,\Gamma })\frac{m_{\Gamma }}{%
1-n_{f\sigma }^{0}}] \\
&&+\sum_{\sigma ^{\prime }}\varepsilon _{\sigma ^{\prime }}\sum_{\Gamma \ni
\sigma ^{\prime }}A_{\sigma ^{\prime },\Gamma }[A_{\sigma ,\Gamma }\frac{%
m_{\Gamma }}{n_{f\sigma }^{0}}-(1-A_{\sigma ,\Gamma })\frac{m_{\Gamma }}{%
1-n_{f\sigma }^{0}}] \\
&=&[A_{\sigma ,\Gamma }\frac{m_{\Gamma }}{n_{f\sigma }^{0}}-(1-A_{\sigma
,\Gamma })\frac{m_{\Gamma }}{1-n_{f\sigma }^{0}}]S_{\Gamma }
\end{eqnarray*}

\bigskip

thus%
\begin{equation*}
\langle +k_{1}\sigma \left\vert \hat{H}_{local}|+k_{2}\sigma \right\rangle
=\langle 0|\hat{c}_{k_{1}\sigma }\hat{c}_{k_{2}\sigma }^{+}\left\vert
0\right\rangle S_{1}+\langle 0|\hat{c}_{k_{1}\sigma }\hat{f}_{\sigma }^{\dag
}\left\vert 0\right\rangle \langle 0|\hat{f}_{\sigma }\hat{c}_{k_{2}\sigma
}^{+}\left\vert 0\right\rangle S_{25}(\sigma )
\end{equation*}

\begin{equation*}
\langle +k\sigma \left\vert \hat{H}_{local}|UHB\right\rangle =\frac{1}{\sqrt{%
n_{f\sigma }^{0}(1-n_{f\sigma }^{0})}}\langle 0|\hat{c}_{k\sigma }\hat{f}%
_{\sigma }^{\dag }\left\vert 0\right\rangle S_{2}(\sigma )
\end{equation*}

\begin{equation*}
\langle +k\sigma \left\vert \hat{H}_{local}|QE\right\rangle =\frac{1}{%
n_{f\sigma }^{0}}\langle 0|\hat{c}_{k\sigma }\hat{f}_{\sigma }^{\dag
}\left\vert 0\right\rangle S_{3}(\sigma )
\end{equation*}

\begin{equation*}
\langle UHB|\hat{H}_{local}|UHB\rangle =S_{4}(\sigma )
\end{equation*}

\begin{equation*}
\langle UHB|\hat{H}_{local}|QE\rangle =\frac{\sqrt{1-n_{f\sigma }^{0}}}{%
\sqrt{n_{f\sigma }^{0}}}S_{2}(\sigma )
\end{equation*}

\begin{equation*}
\langle QE|\hat{H}_{local}|QE\rangle =\frac{1-n_{f\sigma }^{0}}{n_{f\sigma
}^{0}}S_{3}(\sigma )
\end{equation*}

\begin{equation*}
\langle -k_{1}\sigma \left\vert \hat{H}_{local}|-k_{2}\sigma \right\rangle
=\langle 0|\hat{c}_{k_{1}\sigma }^{+}\hat{c}_{k_{2}\sigma }\left\vert
0\right\rangle S_{1}-\langle 0|\hat{c}_{k_{1}\sigma }^{+}\hat{f}_{\sigma
}\left\vert 0\right\rangle \langle 0|\hat{f}_{\sigma }^{\dag }\hat{c}%
_{k_{2}\sigma }\left\vert 0\right\rangle S_{25}(\sigma )
\end{equation*}

\begin{equation*}
\langle -k\sigma \left\vert \hat{H}_{local}|LHB\right\rangle =\frac{1}{\sqrt{%
n_{f\sigma }^{0}(1-n_{f\sigma }^{0})}}\langle 0|\hat{c}_{k\sigma }^{+}\hat{f}%
_{\sigma }\left\vert 0\right\rangle S_{5}(\sigma )
\end{equation*}

\begin{equation*}
\langle -k\sigma \left\vert \hat{H}_{local}|QH\right\rangle =\frac{1}{%
1-n_{f\sigma }^{0}}\langle 0|\hat{c}_{k\sigma }^{+}\hat{f}_{\sigma
}\left\vert 0\right\rangle S_{6}(\sigma )
\end{equation*}

\begin{equation*}
\langle LHB|\hat{H}_{local}|LHB\rangle =S_{7}(\sigma )
\end{equation*}

\begin{equation*}
\langle LHB|\hat{H}_{local}|QH\rangle =\frac{\sqrt{n_{f\sigma }^{0}}}{\sqrt{%
1-n_{f\sigma }^{0}}}S_{5}(\sigma )
\end{equation*}

\begin{equation*}
\langle QH|\hat{H}_{local}|QH\rangle =\frac{n_{f\sigma }^{0}}{1-n_{f\sigma
}^{0}}S_{6}(\sigma )
\end{equation*}

\bigskip

\subsubsection{H\_V$\protect\bigskip $}

\begin{eqnarray*}
\langle +k_{1}\sigma | \hat{H}_{V}|+k_{2}\sigma \rangle &=&\sum_{k^{\prime
}\sigma ^{\prime }}V_{k^{\prime }\sigma ^{\prime }}[\delta _{\sigma \sigma
^{\prime }}z_{\sigma }(\langle 0|\hat{c}_{k_{1}\sigma }\hat{c}_{k^{\prime
}\sigma }^{+}\hat{f}_{\sigma }\hat{c}_{k_{2}\sigma }^{+}\left\vert
0\right\rangle +\langle 0|\hat{c}_{k_{1}\sigma }\hat{f}_{\sigma }^{\dag }%
\hat{c}_{k^{\prime }\sigma }\hat{c}_{k_{2}\sigma }^{+}\left\vert
0\right\rangle ) \\
&&+(1-\delta _{\sigma \sigma ^{\prime }})(\langle 0|\hat{c}_{k^{\prime
}\sigma ^{\prime }}^{+}\hat{f}_{\sigma ^{\prime }}\left\vert 0\right\rangle
+\langle 0|\hat{f}_{\sigma ^{\prime }}^{\dag }\hat{c}_{k^{\prime }\sigma
^{\prime }}\left\vert 0\right\rangle ) \\
&&\times (x_{\sigma \sigma ^{\prime }}\langle 0|\hat{c}_{k_{1}\sigma }\hat{c}%
_{k_{2}\sigma }^{+}\hat{f}_{\sigma }^{\dag }\hat{f}_{\sigma }\left\vert
0\right\rangle +v_{\sigma \sigma ^{\prime }}\langle 0|\hat{c}_{k_{1}\sigma }%
\hat{c}_{k_{2}\sigma }^{+}\hat{f}_{\sigma }\hat{f}_{\sigma }^{\dag
}\left\vert 0\right\rangle )]
\end{eqnarray*}

\begin{eqnarray*}
\langle +k\sigma \left\vert \hat{H}_{V}|UHB\right\rangle &=&\sum_{k^{\prime
}\sigma ^{\prime }}V_{k^{\prime }\sigma ^{\prime }}[\delta _{\sigma \sigma
^{\prime }}\langle 0|\hat{c}_{k\sigma }\hat{c}_{k^{\prime }\sigma }^{+}\hat{f%
}_{\sigma }\hat{f}_{\sigma }^{\dag }\left\vert 0\right\rangle \\
&&+(1-\delta _{\sigma \sigma ^{\prime }})\langle 0|\hat{c}_{k\sigma }\hat{f}%
_{\sigma }^{\dag }\left\vert 0\right\rangle (y_{\sigma \sigma ^{\prime
}}\langle 0|\hat{c}_{k^{\prime }\sigma ^{\prime }}^{+}\hat{f}_{\sigma
^{\prime }}\left\vert 0\right\rangle +w_{\sigma \sigma ^{\prime }}\langle 0|%
\hat{f}_{\sigma ^{\prime }}^{\dag }\hat{c}_{k^{\prime }\sigma ^{\prime
}}\left\vert 0\right\rangle )]
\end{eqnarray*}

\begin{eqnarray*}
\langle +k\sigma \left\vert \hat{H}_{V}|QE\right\rangle &=&\sum_{k^{\prime
}\sigma ^{\prime }}V_{k^{\prime }\sigma ^{\prime }}[\delta _{\sigma \sigma
^{\prime }}z_{\sigma }\langle 0|\hat{c}_{k\sigma }\hat{c}_{k^{\prime }\sigma
}^{+}\hat{f}_{\sigma }\hat{f}_{\sigma }^{\dag }\left\vert 0\right\rangle \\
&&++(1-\delta _{\sigma \sigma ^{\prime }})x_{\sigma \sigma ^{\prime
}}\langle 0|c_{k\sigma }f_{\sigma }^{\dag }\left\vert 0\right\rangle
(\langle 0|c_{k^{\prime }\sigma ^{\prime }}^{+}f_{\sigma ^{\prime
}}\left\vert 0\right\rangle +\langle 0|f_{\sigma ^{\prime }}^{\dag
}c_{k^{\prime }\sigma ^{\prime }}\left\vert 0\right\rangle )]
\end{eqnarray*}

\begin{equation*}
\langle UHB|\hat{H}_{V}|UHB\rangle =\sum_{k^{\prime }\sigma ^{\prime
}}V_{k^{\prime }\sigma ^{\prime }}(1-\delta _{\sigma \sigma ^{\prime
}})v_{\sigma \sigma ^{\prime }}(1-n_{f\sigma }^{0})(\langle 0|\hat{c}%
_{k^{\prime }\sigma ^{\prime }}^{+}\hat{f}_{\sigma ^{\prime }}\left\vert
0\right\rangle +\langle 0|\hat{f}_{\sigma ^{\prime }}^{\dag }\hat{c}%
_{k^{\prime }\sigma ^{\prime }}\left\vert 0\right\rangle )
\end{equation*}

\begin{equation*}
\langle UHB|\hat{H}_{V}|QE\rangle =\sum_{k^{\prime }\sigma ^{\prime
}}V_{k^{\prime }\sigma ^{\prime }}(1-\delta _{\sigma \sigma ^{\prime
}})(1-n_{f\sigma }^{0})(w_{\sigma \sigma ^{\prime }}\langle 0|c_{k^{\prime
}\sigma ^{\prime }}^{+}f_{\sigma ^{\prime }}\left\vert 0\right\rangle
+y_{\sigma \sigma ^{\prime }}\langle 0|f_{\sigma ^{\prime }}^{\dag
}c_{k^{\prime }\sigma ^{\prime }}\left\vert 0\right\rangle )
\end{equation*}

\begin{equation*}
\langle QE|H_{V}|QE\rangle =\sum_{k^{\prime }\sigma ^{\prime }}V_{k^{\prime
}\sigma ^{\prime }}(1-\delta _{\sigma \sigma ^{\prime }})x_{\sigma \sigma
^{\prime }}(1-n_{f\sigma }^{0})(\langle 0|c_{k^{\prime }\sigma ^{\prime
}}^{+}f_{\sigma ^{\prime }}\left\vert 0\right\rangle +\langle 0|f_{\sigma
^{\prime }}^{\dag }c_{k^{\prime }\sigma ^{\prime }}\left\vert 0\right\rangle
)
\end{equation*}

\begin{eqnarray*}
\langle -k_{1}\sigma \left\vert \hat{H}_{V}|-k_{2}\sigma \right\rangle
&=&\sum_{k^{\prime }\sigma ^{\prime }}V_{k^{\prime }\sigma ^{\prime
}}[\delta _{\sigma \sigma ^{\prime }}z_{\sigma }(\langle 0|\hat{c}%
_{k_{1}\sigma }^{+}\hat{c}_{k^{\prime }\sigma }^{+}\hat{f}_{\sigma }\hat{c}%
_{k_{2}\sigma }\left\vert 0\right\rangle +\langle 0|\hat{c}_{k_{1}\sigma
}^{+}\hat{f}_{\sigma }^{\dag }\hat{c}_{k^{\prime }\sigma }\hat{c}%
_{k_{2}\sigma }\left\vert 0\right\rangle ) \\
&&+(1-\delta _{\sigma \sigma ^{\prime }})(\langle 0|\hat{c}_{k^{\prime
}\sigma ^{\prime }}^{+}\hat{f}_{\sigma ^{\prime }}\left\vert 0\right\rangle
+\langle 0|\hat{f}_{\sigma ^{\prime }}^{\dag }\hat{c}_{k^{\prime }\sigma
^{\prime }}\left\vert 0\right\rangle ) \\
&&(x_{\sigma \sigma ^{\prime }}\langle 0|\hat{c}_{k_{1}\sigma }^{+}\hat{c}%
_{k_{2}\sigma }\hat{f}_{\sigma }^{\dag }\hat{f}_{\sigma }\left\vert
0\right\rangle +v_{\sigma \sigma ^{\prime }}\langle 0|\hat{c}_{k_{1}\sigma
}^{+}\hat{c}_{k_{2}\sigma }\hat{f}_{\sigma }\hat{f}_{\sigma }^{\dag
}\left\vert 0\right\rangle )]
\end{eqnarray*}

\begin{eqnarray*}
\langle -k\sigma \left\vert \hat{H}_{V}|LHB\right\rangle &=&\sum_{k^{\prime
}\sigma ^{\prime }}V_{k^{\prime }\sigma ^{\prime }}[\delta _{\sigma \sigma
^{\prime }}\langle 0|\hat{c}_{k\sigma }^{+}\hat{f}_{\sigma }^{\dag }\hat{c}%
_{k^{\prime }\sigma }\hat{f}_{\sigma }\left\vert 0\right\rangle \\
&&+(1-\delta _{\sigma \sigma ^{\prime }})\langle 0|\hat{c}_{k\sigma }^{+}%
\hat{f}_{\sigma }\left\vert 0\right\rangle (w_{\sigma \sigma ^{\prime
}}\langle 0|\hat{c}_{k^{\prime }\sigma ^{\prime }}^{+}\hat{f}_{\sigma
^{\prime }}\left\vert 0\right\rangle +y_{\sigma \sigma ^{\prime }}\langle 0|%
\hat{f}_{\sigma ^{\prime }}^{\dag }\hat{c}_{k^{\prime }\sigma ^{\prime
}}\left\vert 0\right\rangle )]
\end{eqnarray*}%
\qquad \qquad

\begin{eqnarray*}
\langle -k\sigma \left\vert H_{V}|QH\right\rangle &=&\sum_{k^{\prime }\sigma
^{\prime }}V_{k^{\prime }\sigma ^{\prime }}[\delta _{\sigma \sigma ^{\prime
}}z_{\sigma }\langle 0|c_{k\sigma }^{+}f_{\sigma }^{\dag }c_{k^{\prime
}\sigma }f_{\sigma }\left\vert 0\right\rangle \\
&&+(1-\delta _{\sigma \sigma ^{\prime }})v_{\sigma \sigma ^{\prime }}\langle
0|c_{k\sigma }^{+}f_{\sigma }\left\vert 0\right\rangle (\langle
0|c_{k^{\prime }\sigma ^{\prime }}^{+}f_{\sigma ^{\prime }}\left\vert
0\right\rangle +\langle 0|f_{\sigma ^{\prime }}^{\dag }c_{k^{\prime }\sigma
^{\prime }}\left\vert 0\right\rangle )]
\end{eqnarray*}

\begin{equation*}
\langle LHB|\hat{H}_{V}|LHB\rangle =\sum_{k^{\prime }\sigma ^{\prime
}}V_{k^{\prime }\sigma ^{\prime }}(1-\delta _{\sigma \sigma ^{\prime
}})x_{\sigma \sigma ^{\prime }}n_{f\sigma }^{0}(\langle 0|\hat{c}_{k^{\prime
}\sigma ^{\prime }}^{+}\hat{f}_{\sigma ^{\prime }}\left\vert 0\right\rangle
+\langle 0|\hat{f}_{\sigma ^{\prime }}^{\dag }\hat{c}_{k^{\prime }\sigma
^{\prime }}\left\vert 0\right\rangle )
\end{equation*}

\begin{equation*}
\langle LHB|H_{V}|QH\rangle =\sum_{k^{\prime }\sigma ^{\prime }}V_{k^{\prime
}\sigma ^{\prime }}(1-\delta _{\sigma \sigma ^{\prime }})n_{f\sigma
}^{0}(y_{\sigma \sigma ^{\prime }}\langle 0|c_{k^{\prime }\sigma ^{\prime
}}^{+}f_{\sigma ^{\prime }}\left\vert 0\right\rangle +w_{\sigma \sigma
^{\prime }}\langle 0|f_{\sigma ^{\prime }}^{\dag }c_{k^{\prime }\sigma
^{\prime }}\left\vert 0\right\rangle )
\end{equation*}

\begin{equation*}
\langle QH\left\vert H_{V}|QH\right\rangle =\sum_{k^{\prime }\sigma ^{\prime
}}V_{k^{\prime }\sigma ^{\prime }}(1-\delta _{\sigma \sigma ^{\prime
}})v_{\sigma \sigma ^{\prime }}n_{f\sigma }^{0}(\langle 0|c_{k^{\prime
}\sigma ^{\prime }}^{+}f_{\sigma ^{\prime }}\left\vert 0\right\rangle
+\langle 0|f_{\sigma ^{\prime }}^{\dag }c_{k^{\prime }\sigma ^{\prime
}}\left\vert 0\right\rangle )
\end{equation*}





\begin{figure}[!ht]
\includegraphics[height=4cm,width=6cm,bb=2 25 290 245]{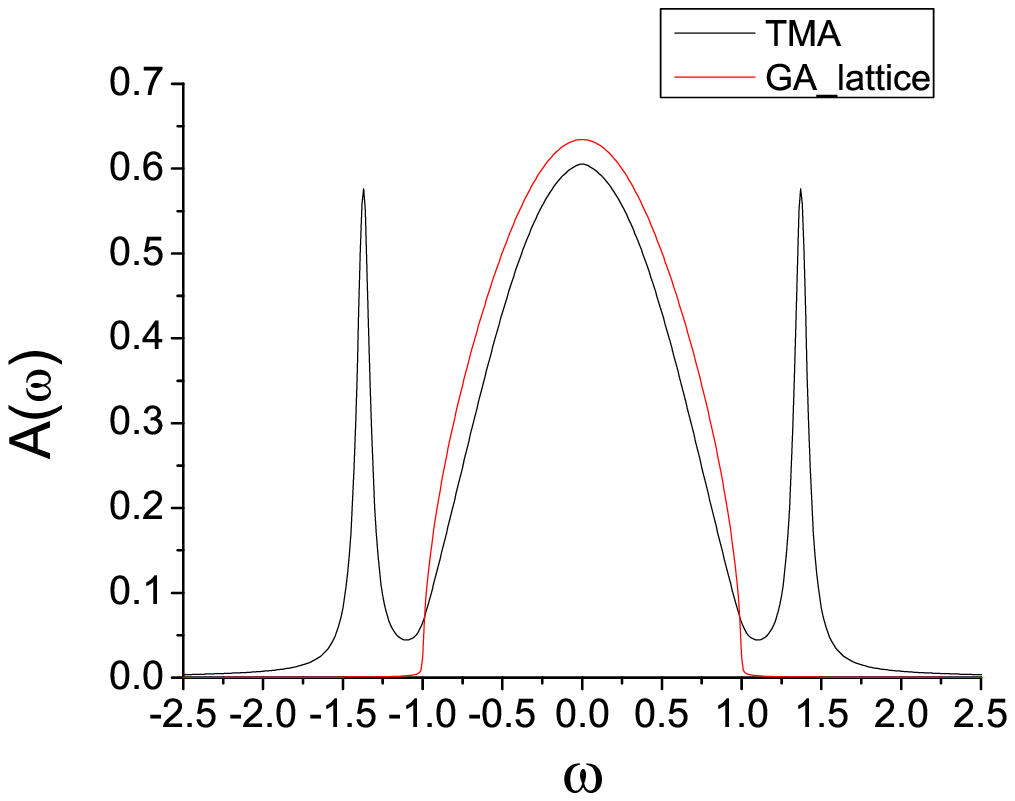}
\caption{The spectral function of electrons on the impurity site for
an single orbital impurity model obtained by TMA and GA lattice with
$U=1$.} \label{GFimp,psVSmf}
\end{figure}

\begin{figure}[!ht]
\includegraphics[height=7.5cm,width=7.5cm,bb=28 40 459
374]{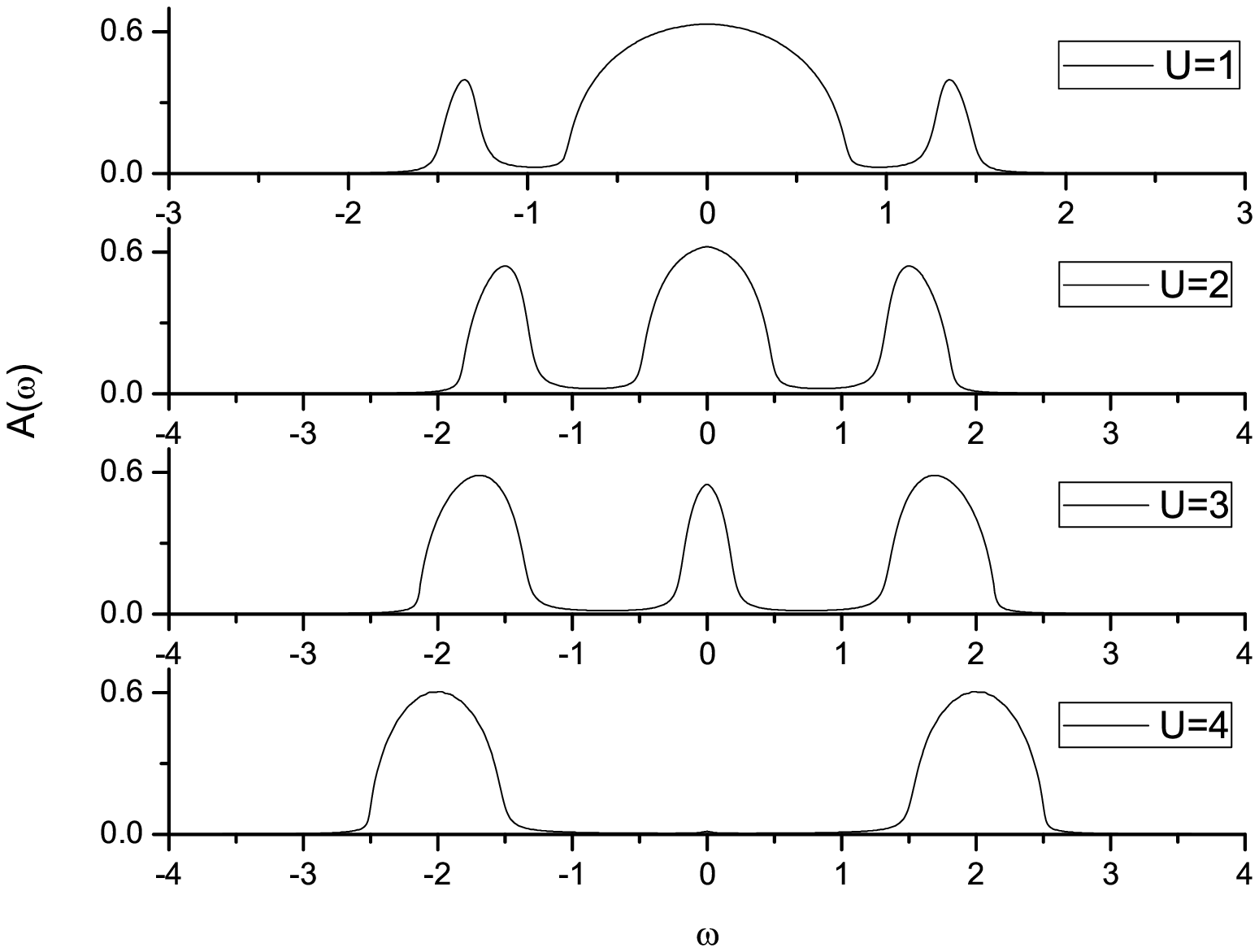} \caption{The density of states (DOS)
obtained by DMFT+TMA for single-band Hubbard model on Bethe lattice
at half filling.} \label{spect,sb,ntot=1}
\end{figure}

\begin{figure}[th]
\includegraphics[height=6cm,width=8cm,bb=19 6 503
380]{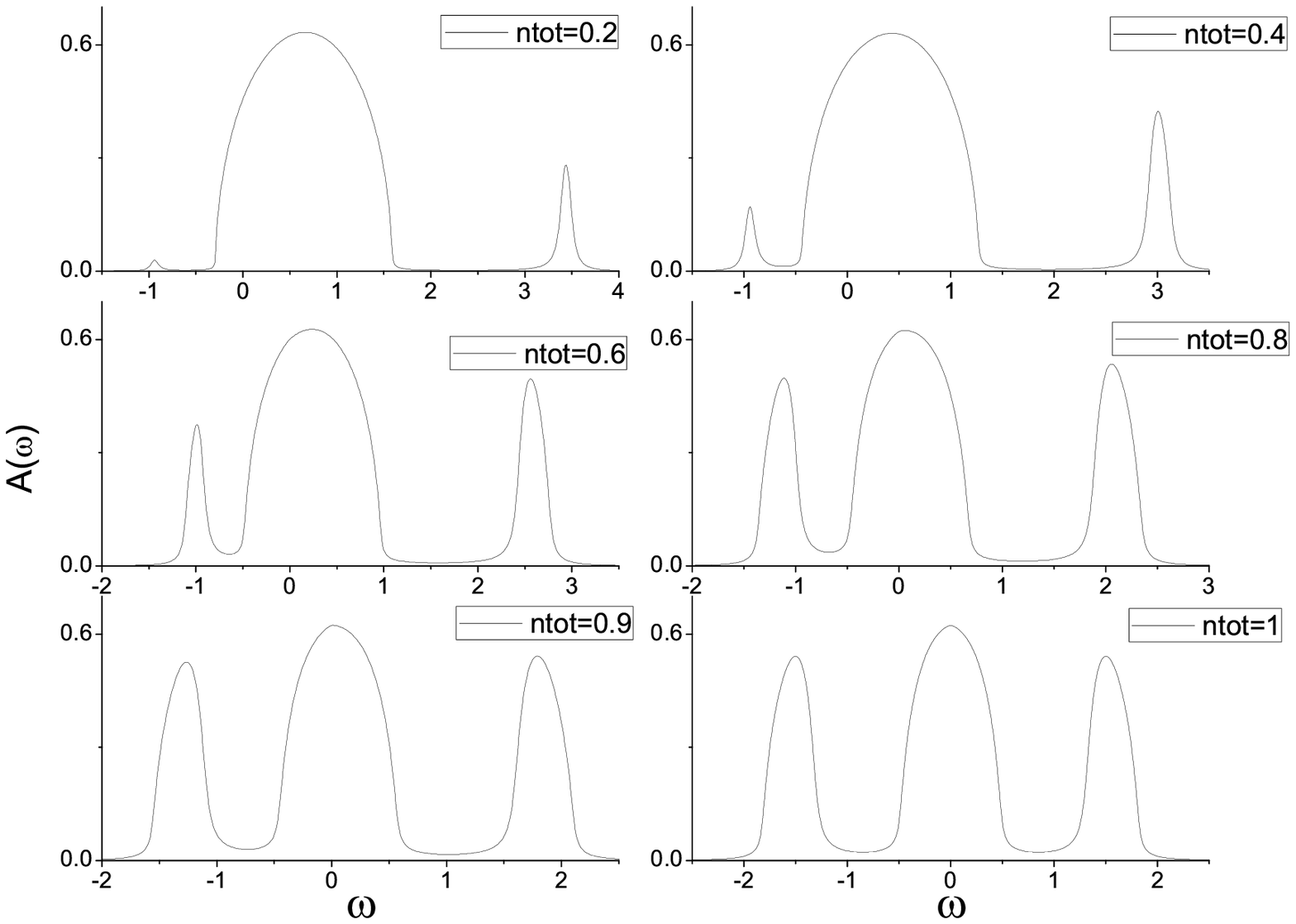} \caption{The density of states (DOS)
obtained by DMFT+TMA of single band Hubbard model under $U=2$ with
different fillings. } \label{U=2,filling varies}
\end{figure}

\begin{figure}[th]
\includegraphics[height=6cm,width=7.5cm,bb=49 16 548 380]{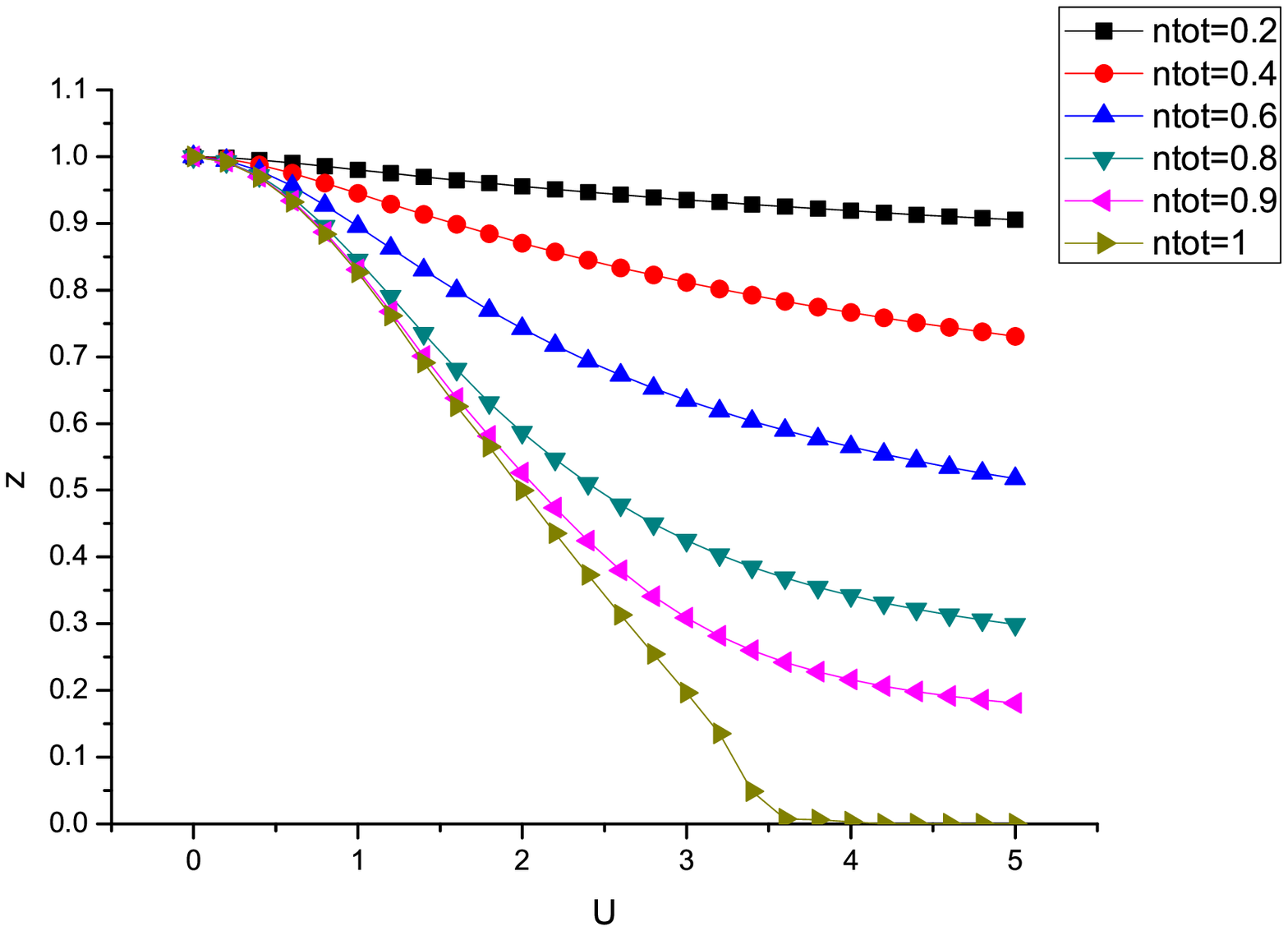}
\caption{Quasi-particle weight $z$ of single band Hubbard model
obtained by DMFT+TMA versus $U$ at different fillings.}
\label{oneband,zU}
\end{figure}

\begin{figure}[th]
\includegraphics[height=7cm,width=8cm,bb=26 20 269
249]{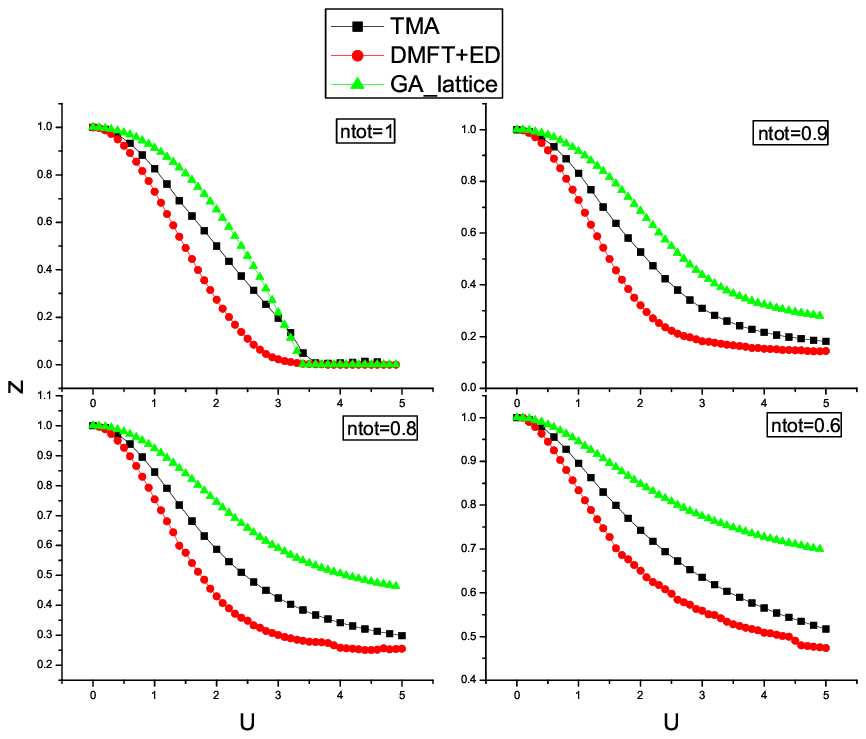} \caption{Comparison of Quasi-particle
weight $z$ for the single band Hubbard model obtained by DMFT+TMA,
GA lattice and DMFT+ED.} \label{1band,z_compare}
\end{figure}

\begin{figure}[th]
\includegraphics[height=6cm,width=6cm,bb=36 33 464 403]{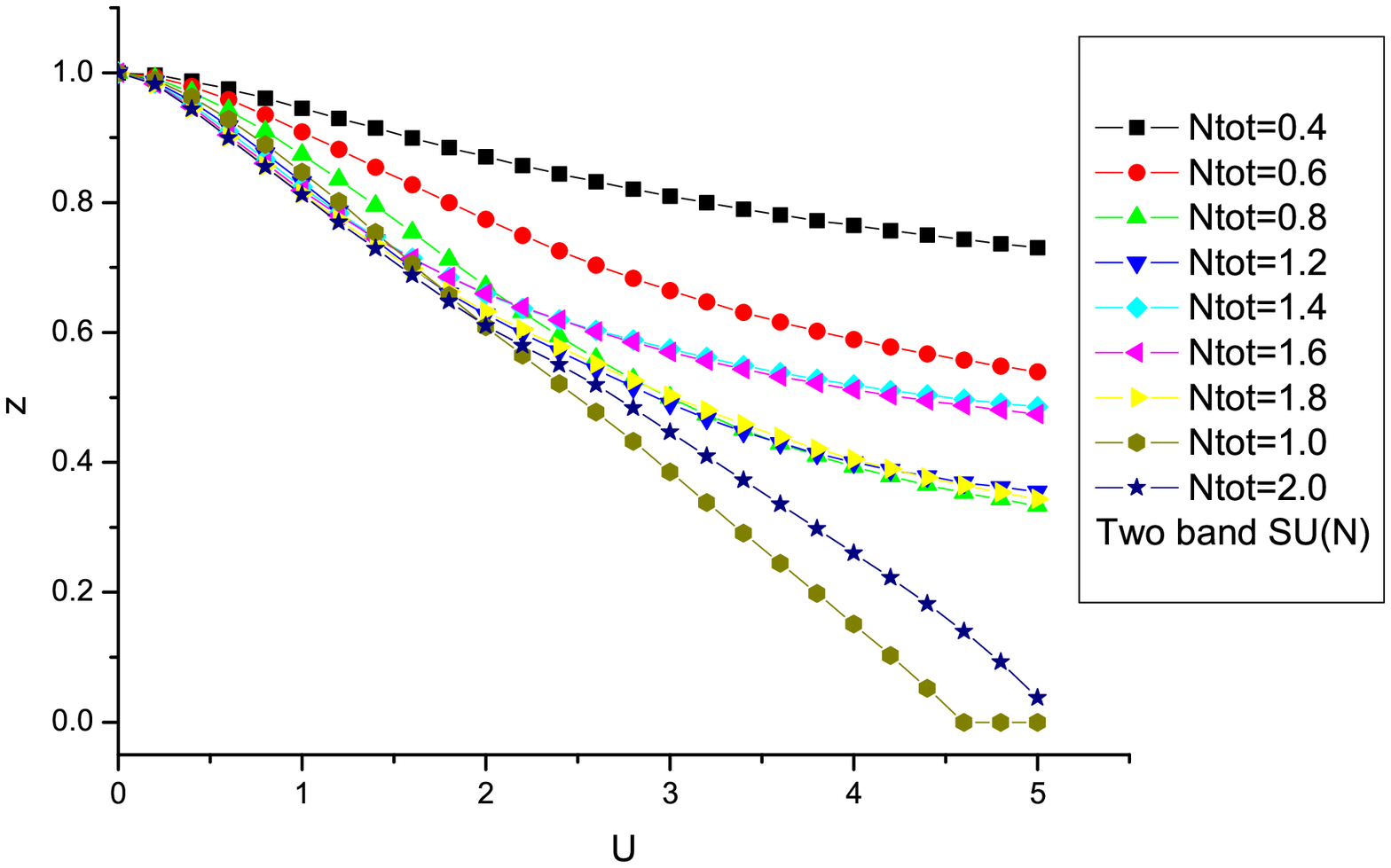}
\caption{Quasi-particle weight $z$ as the function of $U$ for the
two-band Hubbard model with SU(N) symmetry obtained by DMFT+TMA.}
\label{twoband,zU}
\end{figure}

\begin{figure}[th]
\includegraphics[height=10.5cm,width=12cm,bb=26 17 276
253]{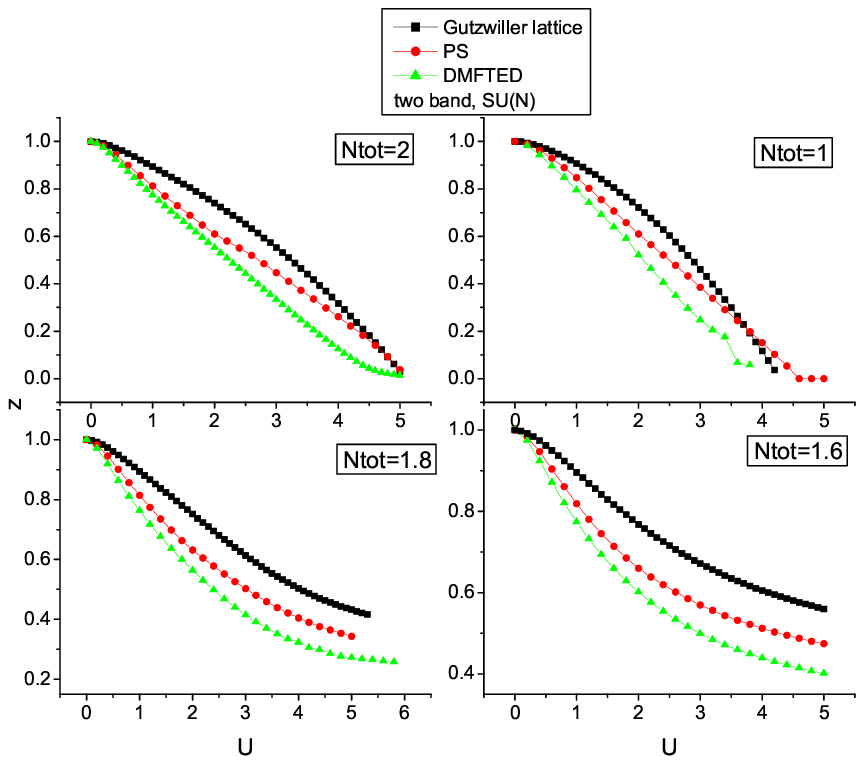} \caption{Comparison of quasi-particle
weight $z$ for the two band Hubbard model with SU(N) symmetry at
different fillings obtained by DMFT+TMA, GA lattice and DMFT+ED.}
\label{twoband,zcompare}
\end{figure}

\begin{figure}[th]
\includegraphics[height=6cm,width=8cm,bb=30 18 294
224]{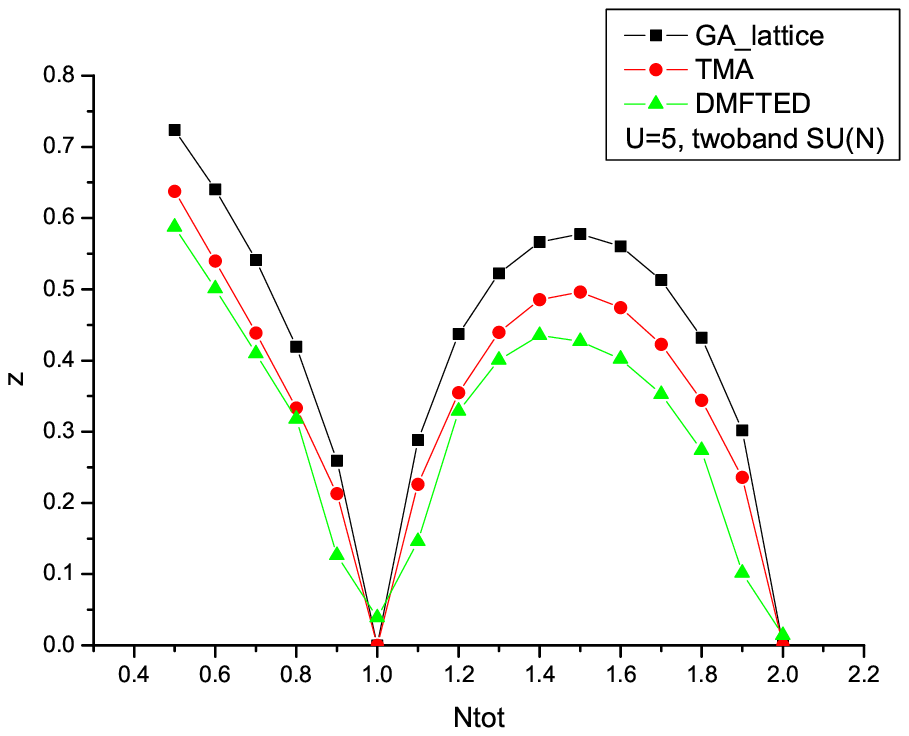} \caption{Comparison of quasi-particle
weight $z$ as the function of total number of particles for the
two-band Hubbard model with SU(N) symmetry at $U=5$ obtained by
DMFT+TMA, GA lattice and DMFT+ED.} \label{twoband,U=5,zN}
\end{figure}

\begin{figure}[th]
\includegraphics[height=6cm,width=7cm,bb=29 14 239
231]{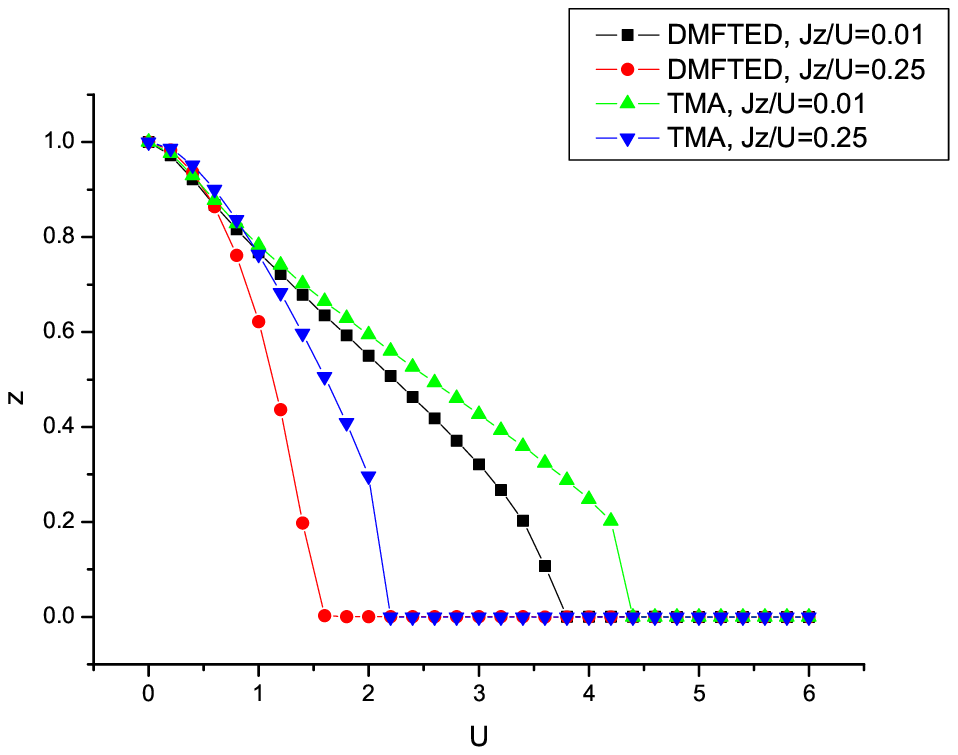} \caption{Comparison of
quasi-particle weight $z$ for the two-degenerate-band Hubbard model
with longitudinal Hund's coupling $J_z$ obtained by DMFT+TMA and
DMFT+ED at different $J_z/U$.} \label{JZoverU,comparewithED}
\end{figure}

\begin{figure}[th]
\includegraphics[height=9cm,width=12cm,bb=36 28 464
403]{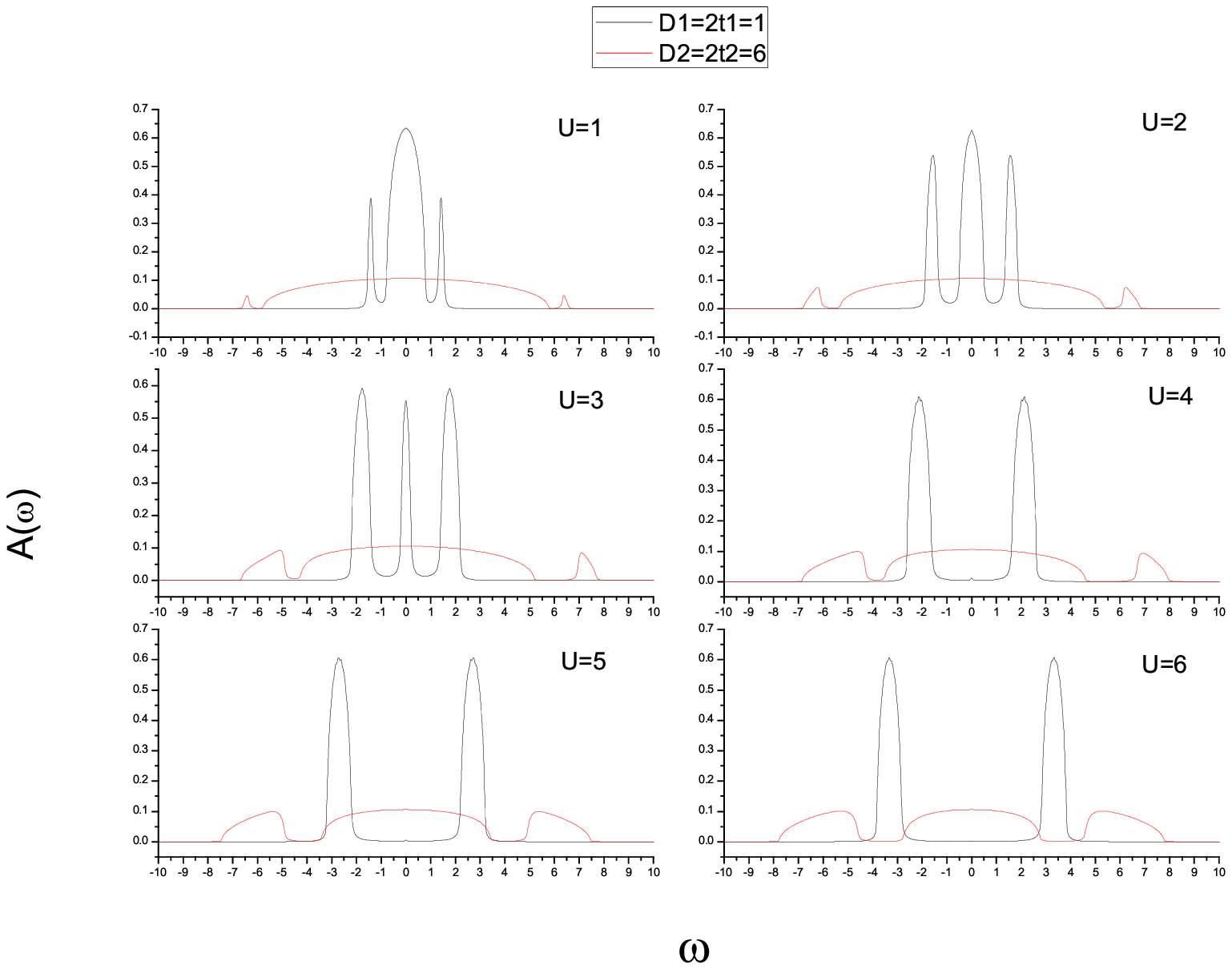} \caption{The spectral
functions obtained by DMFT+TMA for two-nondegenerate-band Hubbard
model with band width ratio $1:6$ under different $U$ with
$J_z=0.3U$.} \label{twoband,t1=0.5,t2=3,JZoverU=0.3}
\end{figure}

\begin{figure}[th]
\includegraphics[height=6cm,width=8cm,bb=36 33 464
403]{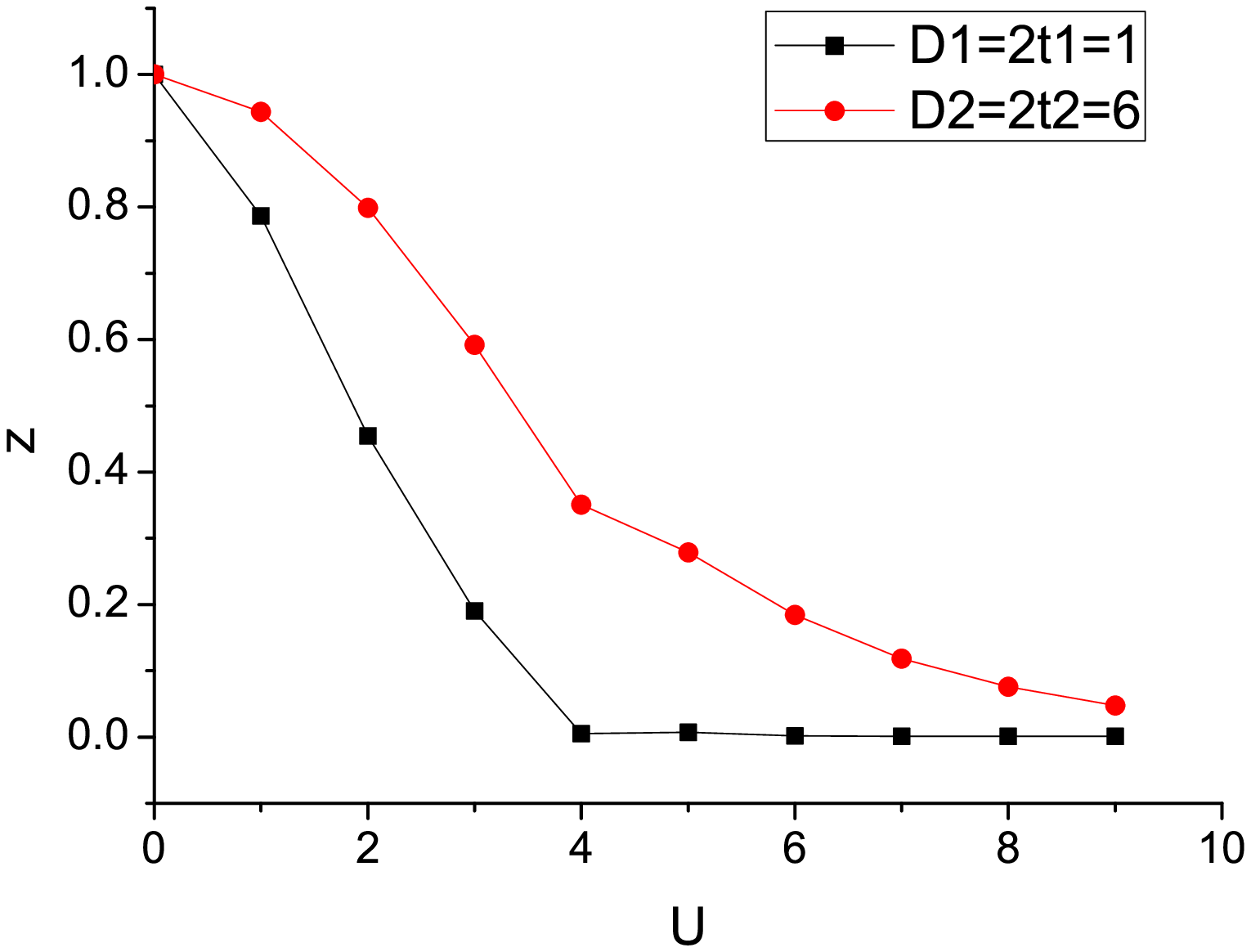} \caption{Quasi-particle
weight $z$ of different bands as the function of $U$ for
two-nondegenerate-band Hubbard model with band width ratio $1:6$
under different $U$ with $J_z=0.3U$.}
\label{twoband,t1=0.5,t2=3,JZoverU=0.3,z}
\end{figure}

\end{document}